\documentclass[11pt,a4paper]{article}
\usepackage[T1]{fontenc}
\usepackage{mathptmx}
\usepackage{graphicx}
\usepackage{bm}
\usepackage{booktabs}
\usepackage{xcolor}
\usepackage{microtype}
\usepackage{mathtools}
\usepackage{amssymb}
\usepackage{array}
\usepackage{float}
\usepackage{multirow}
\usepackage{placeins}
\usepackage{flafter}
\usepackage{setspace}
\usepackage{geometry}
\graphicspath{{./}}
\AtBeginDocument{\color{black}}

\usepackage[
    colorlinks=true,
    linkcolor=black,
    citecolor=black,
    urlcolor=black,
    pdfborder={0 0 0},
    pdftitle={Transport fidelity and domain of validity of compact Gaussian kinetic representations for rarefied flows},
    pdfauthor={Ehsan Roohi}
]{hyperref}
\newcommand{\doilink}[1]{\href{https://doi.org/#1}{DOI}}
\newcommand{\onlinelink}[1]{\href{#1}{Online}}

\begin{document}

\begin{center}
{\LARGE\bfseries Transport fidelity and domain of validity of compact Gaussian kinetic representations for rarefied flows\par}
\vspace{1.1em}
{\large Ehsan Roohi\par}
\vspace{0.45em}
{\normalsize Department of Mechanical and Industrial Engineering, University of Massachusetts Amherst, Amherst, Massachusetts 01003, USA\par}
\vspace{0.25em}
{\normalsize Corresponding author: \href{mailto:roohie@umass.edu}{roohie@umass.edu}\par}
\end{center}
\vspace{0.5em}
\begin{abstract}
Compact representations of rarefied flows must retain nonequilibrium transport information while identifying their range of validity. We investigate a common localized-Gaussian strategy for discrete-velocity-method (DVM) states in monatomic normal shocks and lid-driven cavities. The strategy is specialized to the available kinetic state: a positive phase-space mixture represents shock distributions and regenerates their moment hierarchy by quadrature, whereas a shared-support physical-space map represents 20 cavity fields. Localized support, continuous evaluation, transport fidelity, and coefficient-count accounting therefore provide the common structure across the two benchmarks. For separately fitted Mach-3 and Mach-5 shocks, the method gives sub-percent errors in conserved quantities and approximately 1--2\% errors in transport and higher-order moments. At the same 4608-coefficient budget, the tested multilinear grids produce 89--98\% errors in these nonequilibrium quantities. For both cavity cases, the Gaussian map also outperforms matched bilinear and singular-value-decomposition baselines, reducing maximum errors by factors of approximately 7--24. In the Mach-conditioned tests, a correspondence-preserving local basis reduces the withheld Mach-6 distribution error from $42.86\pm5.40\%$ to $11.45\pm0.94\%$ at fixed storage, while its transport errors remain 30--40\%; a normalized-coordinate guard rejects Mach 12 outside the training range. Independent grid studies confirm that these trends are not dominated by DVM discretization error. The results establish localized Gaussian representations as storage-efficient transport-fidelity maps for fitted kinetic states and provide quantitative acceptance criteria for parametric use.
\end{abstract}
\vspace{0.35em}
\vspace{1.2em}

\section{Introduction}

Rarefied gas simulations contain information at two different levels. The first is the familiar macroscopic level represented by density, mean velocity, and temperature. The second is kinetic: asymmetry and anisotropy of the velocity distribution function (VDF), heat flux, stress, skewness, and closure-sensitive fourth-order content. In a transitional or rarefied flow, agreement in $\rho$, $\bm u$, and $T$ is insufficient because distributions with nearly identical primary fields can transport different momentum and energy. Odd, mixed, and high-order velocity moments measure this departure from local equilibrium and expose errors that are nearly invisible in conserved variables.\cite{Bird1994,Cercignani2000,Sone2007,RoohiStefanov2016,AkhlaghiRoohiStefanov2023}

This hierarchy creates two fluid-mechanical questions for any reduced kinetic description. First, which parts of a normal-shock or wall-bounded distribution are lost first as storage is reduced? Second, does a representation trained over several shock strengths preserve the transport hierarchy at an unseen Mach number, or does increasing nonequilibrium expose a domain boundary? The first question concerns phase-space localization at one state; the second distinguishes reliable parametric interpolation from uncontrolled extrapolation. Both require transport and higher-order moments, rather than a compression ratio alone, as acceptance criteria.

The cost of preserving that information is substantial. A deterministic discrete velocity method (DVM) stores a distribution on the tensor product of a physical grid and a velocity grid. Even a one-dimensional shock with three molecular-velocity coordinates can therefore contain hundreds of millions of scalar values. Large kinetic databases are increasingly used for design studies, uncertainty analysis, reduced models, and operator learning, but raw phase-space arrays are expensive to archive and awkward to compare across grids. A useful representation should reduce storage while retaining a physically meaningful object: it should remain positive when it represents a distribution, be evaluable away from the original storage nodes, and reproduce the same moment hierarchy used to assess the reference solver.

Normal shocks and wall-bounded cavities expose two forms of this same representation problem. When a complete VDF is available, the compact model should preserve a nonnegative phase-space object from which moments can be regenerated. When a database supplies a selected set of wall-transport fields, the compact model should encode those signed outputs with one shared localized support. In both settings, Gaussian kernels allocate a continuous coefficient budget to regions where the kinetic state varies, and fidelity is judged through low-order, transport, and higher moments at matched storage. The two models used below are therefore complementary realizations of one localized-support principle, with the output constraint chosen to match the stored kinetic information.

Gaussian components provide a natural localized basis, but the generic use of Gaussians is not itself new. Mott-Smith represented shock distributions by two Maxwellian populations.\cite{MottSmith1951} Alekseenko, Grandilli, and Wood developed an ultra-sparse isotropic-Gaussian approximation for spatially homogeneous non-continuum distributions.\cite{AlekseenkoGrandilliWood2020} Gaussian-mixture-model compression has also been investigated for plasma distribution functions.\cite{HuPlasmaGMM2025} These studies establish that velocity-space mixtures can represent local or homogeneous kinetic states. The distinct question addressed here is whether one compact continuous representation can resolve the coupled spatial--velocity structure of a shock and preserve transport moments after integration, while a related physical-space representation simultaneously compresses a wall-bounded set of nonequilibrium moment fields.

The same issue has recently emerged in continuum flow compression. Shenoy and Frankel fit localized Gaussian primitives independently to three-dimensional Taylor--Green velocity snapshots and evaluate vorticity, enstrophy, spectra, anisotropy, and scale separation.\cite{ShenoyFrankel2026} Their Gaussian codec is intentionally a snapshot representation: it does not condition one Gaussian model on time and does not test interpolation or extrapolation from one flow stage to another. The present work advances that foundation in five directions. It represents a positive four-dimensional phase-space distribution rather than only a three-dimensional velocity field; regenerates heat flux, stress, and higher kinetic moments by the original DVM quadrature rather than only spatial derivatives; shares one correspondence-preserving conditional model across nine shock strengths; withholds complete kinetic states and tests matched-budget global and local Mach maps; and adds a 20-channel wall-bounded rarefied-flow representation with matched-storage bilinear and singular-value-decomposition (SVD) controls. Gaussian primitives have also become prominent in continuous scene and signal representations,\cite{Kerbl2023,Sitzmann2020} while reduced-order fluid modeling has long relied on proper orthogonal decomposition (POD), SVD, and projection methods.\cite{Sirovich1987,Berkooz1993,Benner2015}

High-dimensional kinetic equations have additionally motivated dynamical low-rank and tensor representations that reduce the cost of evolving phase-space solutions, as well as neural sparse representations that combine continuous spatial models with low-rank velocity factors.\cite{EinkemmerLubich2018,LiNeuralSparse2024} Learned kinetic solvers have also targeted the expensive collision step through data-driven corrections to Bhatnagar--Gross--Krook (BGK) relaxation, accelerated neural collision operators, and structure-preserving relaxation networks.\cite{MillerRobertsBondCyr2022,XiaoFrank2021,XiaoFrank2023} Those approaches primarily reduce kinetic-equation solution cost. The present work addresses a complementary database problem: a converged kinetic state is converted into a positive, continuously queryable representation whose moment errors are evaluated with the same quadrature as the reference solver. This distinction motivates coefficient-matched storage comparisons rather than time-integration speed comparisons.

The work is complementary to operator learning. Deep operator networks (DeepONets), Fourier neural operators, and rarefied-flow surrogates aim to map problem inputs to solutions.\cite{Brunton2020,LuDeepONet2021,LiFNO2021} Within rarefied-gas dynamics, physics-informed neural networks (PINNs) have solved forward and inverse BGK problems and Poiseuille flow, inferred effective viscosity, and learned microflows from direct simulation Monte Carlo (DSMC) data across Knudsen number.\cite{LouMengKarniadakis2021,DeFlorio2022,Tucny2024,Tucny2025} Recent neural surrogates have also addressed micro-step flow, shock waves, cavities, and hypersonic configurations.\cite{RoohiMahdaviDeepONet2026,RoohiShojaSaniPoF2026,RoohiShojaSaniAST2026} A fitted representation instead converts one or several converged kinetic states into a compact differentiable database object. Here fitted-state reconstruction, complete-state holdouts, and range screening are reported as three distinct measures of fidelity. Conditioning on Mach number then tests whether the localized representation can be shared across states while retaining the same transport-based evaluation contract.

\begin{figure}[!tbp]
\centering
\includegraphics[width=\textwidth]{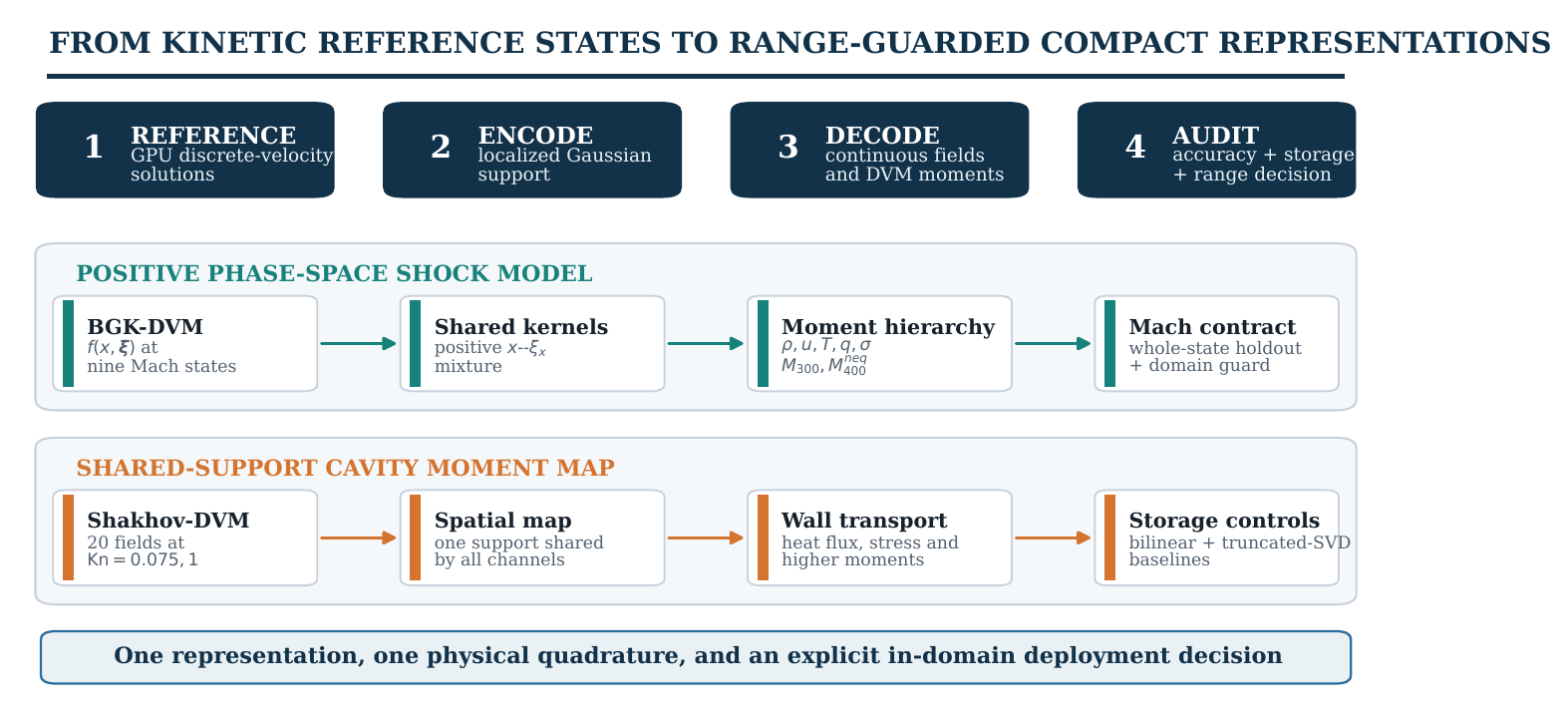}
\caption{Computational workflow. Author-generated Bhatnagar--Gross--Krook (BGK) DVM shock states and Shakhov-DVM cavity states are normalized and represented by either a positive phase-space Gaussian model or a physical-space Gaussian moment map. Shock observables are recovered by the original velocity quadrature. Objective, seed, held-out-station, leave-one-Mach-out, grid-refinement, and matched-storage tests quantify representation, optimization, and reference-grid effects.}
\label{fig:workflow}
\end{figure}

Under this common localized-Gaussian strategy, the study makes six contributions. First, a positive Gaussian mixture is trained directly on $\log f(x,\bm\xi)$ for Mach-3 and Mach-5 shocks, and every reported observable is regenerated by the DVM quadrature. Second, objective, covariance, capacity, seed, and checkpoint tests quantify sensitivity at the configurations examined. Third, interleaved withheld stations compare errors on sampled and unsampled physical locations within one M5 state. Fourth, matched-storage regular-grid, bilinear, and per-field truncated-SVD baselines provide coefficient-count controls. Fifth, wall-bounded maps at $\mathrm{Kn}=0.075$ and 1 are evaluated over all 20 fields and compared through low-order, transport, and higher-moment profiles. Sixth, 24 conditional runs compare global-polynomial and local piecewise-linear Mach dependence over three seeds and complete excluded shock states. Independent grid-refinement studies for Mach 6 and Mach 12 bound the contribution of velocity-grid error. Figure~\ref{fig:workflow} summarizes the workflow.

\section{Kinetic reference problems}

\subsection{BGK and Shakhov models and discrete velocity moments}

All reference fields were generated by the author with graphics-processing-unit (GPU) discrete-velocity solvers and serve as fixed numerical targets for the representation study. The reported Gaussian errors therefore measure fidelity to the stored DVM distributions, fields, and quadrature moments. The normal shocks use the BGK relaxation model,
\begin{equation}
\frac{\partial f}{\partial t}+\bm\xi\cdot\nabla_{\bm x}f=\frac{f^S-f}{\tau},
\label{eq:bgk}
\end{equation}
with $f^S=f^M$ in the shock calculations. Here $f=f(\bm x,\bm\xi,t)$ is the molecular velocity distribution, $\bm x$ is physical position, $\bm\xi$ is molecular velocity, $t$ is time, and $\tau$ is the relaxation time. In the historical nondimensional shock solver used to generate every shock array analyzed here, the upstream relaxation time is the time unit and the cellwise value is the constant
\begin{equation}
\begin{aligned}
\tau_{\mathrm{shock}}&=1,\\
f^{n+1}&=f^M+(f^*-f^M)\exp(-\Delta t/\tau_{\mathrm{shock}})\\
&=f^M+(f^*-f^M)\exp(-\Delta t),
\end{aligned}
\label{eq:exact_relaxation}
\end{equation}
where $f^*$ is the post-transport state. Thus the exact relaxation multiplier in each shock step is $\exp(-\Delta t)$; no temperature-dependent viscosity exponent is applied in the archived shock calculations. The local Maxwellian is
\begin{equation}
 f^M=\frac{\rho}{(2\pi T)^{3/2}}\exp\left(-\frac{|\bm c|^2}{2T}\right),
\label{eq:maxwellian}
\end{equation}
The cavity calculations instead use the Shakhov target
\begin{equation}
 f^S=f^M\left[1+(1-\mathrm{Pr})\frac{\bm c\cdot\bm q}{5pT}
 \left(\frac{|\bm c|^2}{T}-5\right)\right],
\label{eq:shakhov_target}
\end{equation}
with $\mathrm{Pr}=2/3$. In these expressions $\bm c=\bm\xi-\bm u$, $p=\rho T$, and $\bm q$ is heat flux. The shock value in Eq.~\eqref{eq:exact_relaxation} is directly verified in the archived solver. Thus the shock targets correspond to the constant-$\tau$ Maxwellian BGK model with $\mathrm{Pr}_{\mathrm{BGK}}=1$, while the cavity targets use the Shakhov model with $\mathrm{Pr}=2/3$.\cite{BGK1954,Shakhov1968,Mieussens2000,XuHuang2010}

Equation~\eqref{eq:bgk} defines the common transport--relaxation balance used by both reference solvers. Equation~\eqref{eq:exact_relaxation} records the actual constant relaxation coefficient used for every shock state in this study. Equation~\eqref{eq:maxwellian} supplies the equilibrium target for the shock calculations, whereas the heat-flux correction in Eq.~\eqref{eq:shakhov_target} changes the cavity relaxation target so that the specified Shakhov Prandtl number, rather than the BGK value of unity, controls energy transport in the supplied reference fields.

Let $\{(\bm\xi_n,w_n)\}_{n=1}^{N_v}$ denote the discrete velocity nodes and quadrature weights, and let $f_n=f(\bm\xi_n)$. The macroscopic fields are
\begin{align}
\rho &= \sum_{n=1}^{N_v} f_nw_n, &
\rho u_i &= \sum_{n=1}^{N_v}\xi_{i,n}f_nw_n, \label{eq:macro1}\\
3\rho T &= \sum_{n=1}^{N_v}|\bm c_n|^2f_nw_n, &
\bm c_n&=\bm\xi_n-\bm u. \label{eq:macro2}
\end{align}
Equation~\eqref{eq:macro2} closes the low-order quadrature by defining both temperature and the peculiar velocity relative to the recovered bulk motion; consequently, every higher moment below is evaluated in the same local rest frame.
The transport and higher-order observables used throughout the paper are
\begin{align}
\sigma_{ij}&=\sum_{n=1}^{N_v}c_{i,n}c_{j,n}f_nw_n-\rho T\delta_{ij}, \label{eq:stress}\\
q_i&=\frac12\sum_{n=1}^{N_v}|\bm c_n|^2c_{i,n}f_nw_n, \label{eq:heatflux}\\
M_{300}^{\mathrm{neq}}&=\sum_{n=1}^{N_v}c_{x,n}^3f_nw_n, \label{eq:m300}\\
M_{400}^{\mathrm{neq}}&=\sum_{n=1}^{N_v}c_{x,n}^4f_nw_n-3\rho T^2. \label{eq:m400}
\end{align}
Here $i$ and $j$ denote Cartesian components and $\delta_{ij}$ is the Kronecker delta. Equation~\eqref{eq:stress} measures the anisotropic part of the pressure tensor, while Eq.~\eqref{eq:heatflux} weights molecular energy by directed peculiar velocity and therefore diagnoses nonequilibrium energy transport. The odd moment in Eq.~\eqref{eq:m300} detects streamwise distribution skewness. The fourth-order quantity in Eq.~\eqref{eq:m400} is the nonequilibrium deviation rather than the raw fourth moment; subtracting $3\rho T^2$ prevents the equilibrium background from masking reconstruction error.

\subsection{Normal shocks}

The first benchmark consists of stationary monatomic normal shocks. Single-state architecture and matched-storage tests use upstream Mach numbers 3 (M3) and 5 (M5). Shock variables are nondimensionalized by the upstream state, denoted by subscript 1. The plotted coordinate is $x/\lambda_1=x^*/\mathrm{Kn}_{\mathrm{eff}}$, where $x^*\in[-1/2,1/2]$ is the solver coordinate and $\lambda_1$ is the upstream mean free path. Rankine--Hugoniot states are imposed at the domain boundaries. The M3 reference contains 1800 physical stations and 64,009 velocity nodes arranged as $121\times23\times23$; the M5 reference contains 2200 stations and 102,789 nodes arranged as $141\times27\times27$. Their raw phase-space arrays contain approximately $1.15\times10^8$ and $2.26\times10^8$ scalar values.

The two Mach numbers provide different profile scales. The M3 transition is narrower, whereas the M5 profiles span larger downstream-temperature and fourth-moment ranges. The error audit therefore includes both low-order Rankine--Hugoniot-scale variation and localized nonequilibrium profiles; the corresponding reconstructions are presented with the single-state results.

The parametric audit uses nine independently converged BGK-DVM shocks at $M=1.5$, 2, 2.5, 3, 4, 5, 6, 8, and 12. Their physical grids contain 1400--4200 stations and their velocity grids contain 35,017--270,725 nodes. The resulting nine raw arrays contain $2.59\times10^9$ phase values, or 10.36 GB in float32. Each leave-one-Mach-out model is trained on eight complete states and evaluated on all nine. M3 and Mach 6 (M6) are interpolation tests because the training data bracket them; Mach 12 (M12) is a deliberately severe extrapolation test because its removal changes the training interval from $[1.5,12]$ to $[1.5,8]$.

\subsection{Lid-driven cavities}

The second benchmark is a square lid-driven cavity with isothermal fully diffuse walls, lid speed $U_{\mathrm{lid}}=0.1$, wall temperature $T_w=1$, and Knudsen numbers $\mathrm{Kn}=0.075$ and 1. The physical grid contains $65\times65$ points. The velocity domain uses 41 nodes per Cartesian direction over $[-5,5]^3$. The stored state has 20 channels, and the complete DVM configurations are summarized in Table~\ref{tab:dvm_setup}:
\begin{multline}
\bm U=(\rho,u,v,T,q_x,q_y,\Theta_x,\Theta_y,\Theta_z,\\
\sigma_{xx},\sigma_{yy},\sigma_{xy},M_{3x},M_{3y},S_x,S_y,\\
M_{4x},M_{4y},K_x,K_y),
\label{eq:cavity_state}
\end{multline}
Equation~\eqref{eq:cavity_state} fixes the complete 20-channel output contract used in every cavity error and storage comparison. Here $\Theta_i$ are directional temperatures, $M_{3i}$ and $S_i$ are third-order channels, $M_{4i}$ are fourth-order channels, and $K_i$ are closure diagnostics produced by the DVM post-processing. This signed multi-field output and the positive shock VDF instantiate the same localized-Gaussian principle in two state spaces: positivity is enforced for the distribution, while one shared support geometry couples the 20 cavity channels.

The two cavity solutions are not interchangeable tests. At $\mathrm{Kn}=0.075$, wall layers and recirculation coexist with relatively smooth interior fields. At $\mathrm{Kn}=1$, molecular transport spans a much larger fraction of the cavity and the vertical heat flux becomes the most difficult channel. Earlier rarefied-cavity studies provide the physical context for wall transport, thermal edge effects, and high-order fields.\cite{ZhuRoohiEbrahimi2023,RafieenasabRoohiManela2026}

The high-Mach shock rows below use composite three-point Gauss--Legendre (GL3) velocity grids.

\begin{table}[t]
\caption{Reference DVM configurations and numerical verification. The M3 and M5 audits recompute moments from stored distributions. M6 and M12 report the largest cross-grid discrepancy among the listed low-order and nonequilibrium moments.}
\label{tab:dvm_setup}
\centering
\small
\resizebox{\textwidth}{!}{%
\begin{tabular}{lcccccc}
\toprule
Case & model & physical grid & velocity grid & domain & rarefaction & verification\\
\midrule
M3 shock & BGK & $N_x=1800$ & $121\times23\times23$ & $|\xi|\le16$ & $\mathrm{Kn}_{\rm eff}=1/120$ & stored-$f$ $<1.4\times10^{-6}$\\
M5 shock & BGK & $N_x=2200$ & $141\times27\times27$ & $|\xi|\le22$ & $\mathrm{Kn}_{\rm eff}=1/160$ & stored-$f$ $<9.4\times10^{-7}$\\
M6 shock & BGK & $N_x=2880$ & $75\times72\times72$ & composite GL3 & $x/\lambda_1\in[-95,95]$ & medium--fine $<0.12\%$\\
M12 shock & BGK & $N_x=4200$ & $63\times60\times60$ & composite GL3 & $x/\lambda_1\in[-240,240]$ & coarse--medium $<0.97\%$\\
Cavity & Shakhov & $65\times65$ & $41$ per direction & $[-5,5]^3$ & $\mathrm{Kn}=0.075$ & residual $8.20\times10^{-8}$\\
Cavity & Shakhov & $65\times65$ & $41$ per direction & $[-5,5]^3$ & $\mathrm{Kn}=1$ & residual $9.56\times10^{-8}$\\
\bottomrule
\end{tabular}%
}
\end{table}

\subsection{Convergence and internal verification}

The reference states are locked before representation training. As summarized in Table~\ref{tab:dvm_setup}, direct recomputation of $\rho$, $u_x$, $T$, $q_x$, and $\sigma_{xx}$ from the stored M3 and M5 distributions reproduces the archived profiles with maximum relative $L_2$ discrepancies below $1.4\times10^{-6}$ and $9.4\times10^{-7}$, respectively.

High-Mach sensitivity was checked independently by repeating M6 and M12 with the composite GL3 velocity grids and float64 arithmetic. The M6 medium and fine grids use 226,800 and 388,800 velocity nodes; their relative discrepancies are 0.0057--0.0124\% for $\rho$, $u_x$, and $T$ and 0.095--0.114\% for $q_x$, $\sigma_{xx}$, $M_{300}^{\mathrm{neq}}$, and $M_{400}^{\mathrm{neq}}$. The M12 coarse and medium grids use 34,816 and 226,800 velocity nodes; the corresponding discrepancies are 0.055--0.104\% and 0.273--0.961\%. Table~\ref{tab:shock_grid_cert} reports every channel. The converged M6 fine and M12 medium profiles are used for the independent moment re-scoring. An incomplete M12 fine checkpoint is retained only as an audit record and is not used in any reported metric.

\begin{table}[t]
\caption{Independent BGK-DVM grid-refinement discrepancies for the high-Mach references. Entries are relative $L_2$ differences in percent. The maximum difference is 0.11448\% for M6 medium--fine and 0.96139\% for M12 coarse--medium.}
\label{tab:shock_grid_cert}
\centering
\small
\begin{tabular}{lccccccc}
\toprule
Grid pair & $\rho$ & $u_x$ & $T$ & $q_x$ & $\sigma_{xx}$ & $M_{300}^{\rm neq}$ & $M_{400}^{\rm neq}$\\
\midrule
M6 medium--fine & 0.00569 & 0.00828 & 0.01238 & 0.10165 & 0.11448 & 0.09541 & 0.09708\\
M12 coarse--medium & 0.08202 & 0.05513 & 0.10359 & 0.27290 & 0.74370 & 0.65464 & 0.96139\\
\bottomrule
\end{tabular}
\end{table}

The cavity DVM iterations were continued until the largest monitored residual was below $10^{-7}$. Table~\ref{tab:cavity_convergence} reports the terminal values: both cases satisfy the prescribed threshold, and the final maximum residual is controlled by $v$. The representation errors below are evaluated against these terminal reference states.

\begin{table}[t]
\caption{Terminal convergence record for the cavity Shakhov-DVM references.}
\label{tab:cavity_convergence}
\centering
\small
\setlength{\tabcolsep}{3.5pt}
\begin{tabular}{@{}lccc@{}}
\toprule
$\mathrm{Kn}$ & iterations & \shortstack{maximum\\residual} & \shortstack{controlling\\field}\\
\midrule
0.075 & 19,900 & $8.20\times10^{-8}$ & $v$\\
1 & 13,500 & $9.56\times10^{-8}$ & $v$\\
\bottomrule
\end{tabular}
\end{table}

\begin{figure}[!tbp]
\centering
\includegraphics[width=\textwidth]{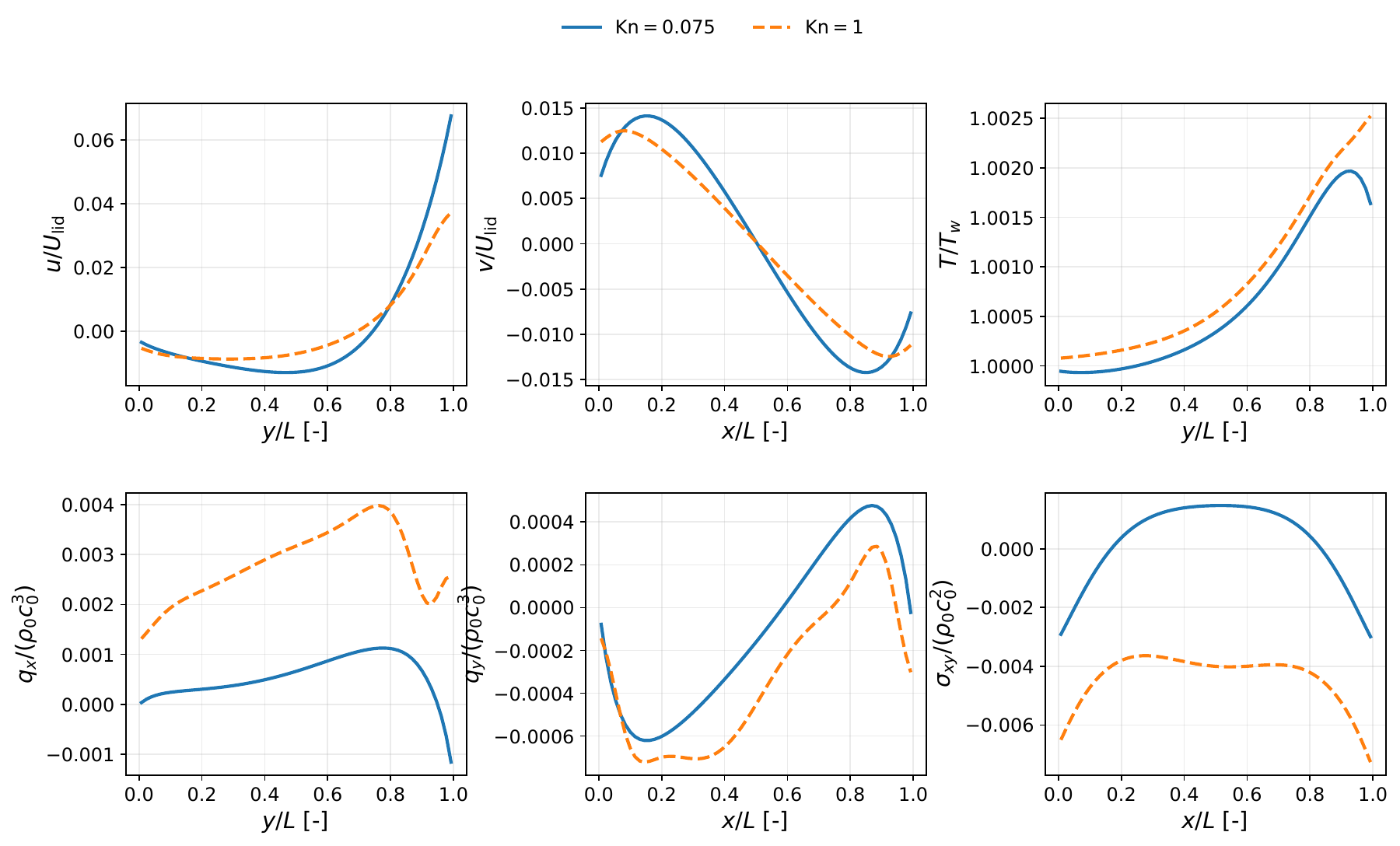}
\caption{DVM and $N=512$ Gaussian centerline profiles at
$\mathrm{Kn}=0.075$ and $1$. Vertical centerlines are shown for
$u$, $T$, and $q_x$, whereas horizontal centerlines are shown for
$v$, $q_y$, and $\sigma_{xy}$. Increasing rarefaction weakens the
lid-driven recirculation, substantially reorganizes the heat-flux
components, shifts the $q_y$ zero crossings, and changes the
horizontal shear-stress profile from a weak positive interior region
to a predominantly negative state. The DVM and Gaussian curves are
shown on identical axes so that these profile differences and the
remaining reconstruction discrepancies can be inspected directly.}
\label{fig:dvm_centerlines}
\end{figure}

\FloatBarrier
Figure~\ref{fig:dvm_centerlines} compares the two converged DVM states and their $N=512$ Gaussian reconstructions. The DVM curves show that the maximum vertical-centerline velocity decreases from 0.0676 at $\mathrm{Kn}=0.075$ to 0.0373 at $\mathrm{Kn}=1$, while the extrema of the horizontal-centerline velocity also decrease. These observations are consistent with the larger wall-slip and weaker collisional coupling expected as rarefaction increases, but the profiles alone are not used to isolate a molecular mechanism.\cite{ZhuRoohiEbrahimi2023,RafieenasabRoohiManela2026}

\FloatBarrier

In Fig.~\ref{fig:dvm_centerlines}, the temperature remains within approximately 0.5\% of the wall value in both cases, whereas the heat-flux profiles change in sign and magnitude. At $\mathrm{Kn}=0.075$, $q_x$ changes sign near the moving lid; at $\mathrm{Kn}=1$, it is positive along the plotted vertical centerline. The horizontal $q_y$ profile crosses zero for both cases, but its branches and crossing locations differ. These plotted differences show that similar temperature levels do not imply similar heat-flux fields; they do not, by themselves, constitute a separate constitutive-law test.

The shear-stress profiles also change qualitatively. At $\mathrm{Kn}=0.075$, $\sigma_{xy}$ has a positive interior region and negative wall-adjacent values, whereas at $\mathrm{Kn}=1$ it is negative along the plotted horizontal centerline. The two DVM states are therefore not simple amplitude rescalings of one another. On the same centerlines, the $N=512$ Gaussian curves closely overlay the DVM curves, including the plotted heat-flux zero crossings and the change in the sign pattern of $\sigma_{xy}$.

\begin{figure}[!tbp]
\centering
\includegraphics[
    width=\textwidth,
    trim=0 0.05cm 0 0.10cm,
    clip
]{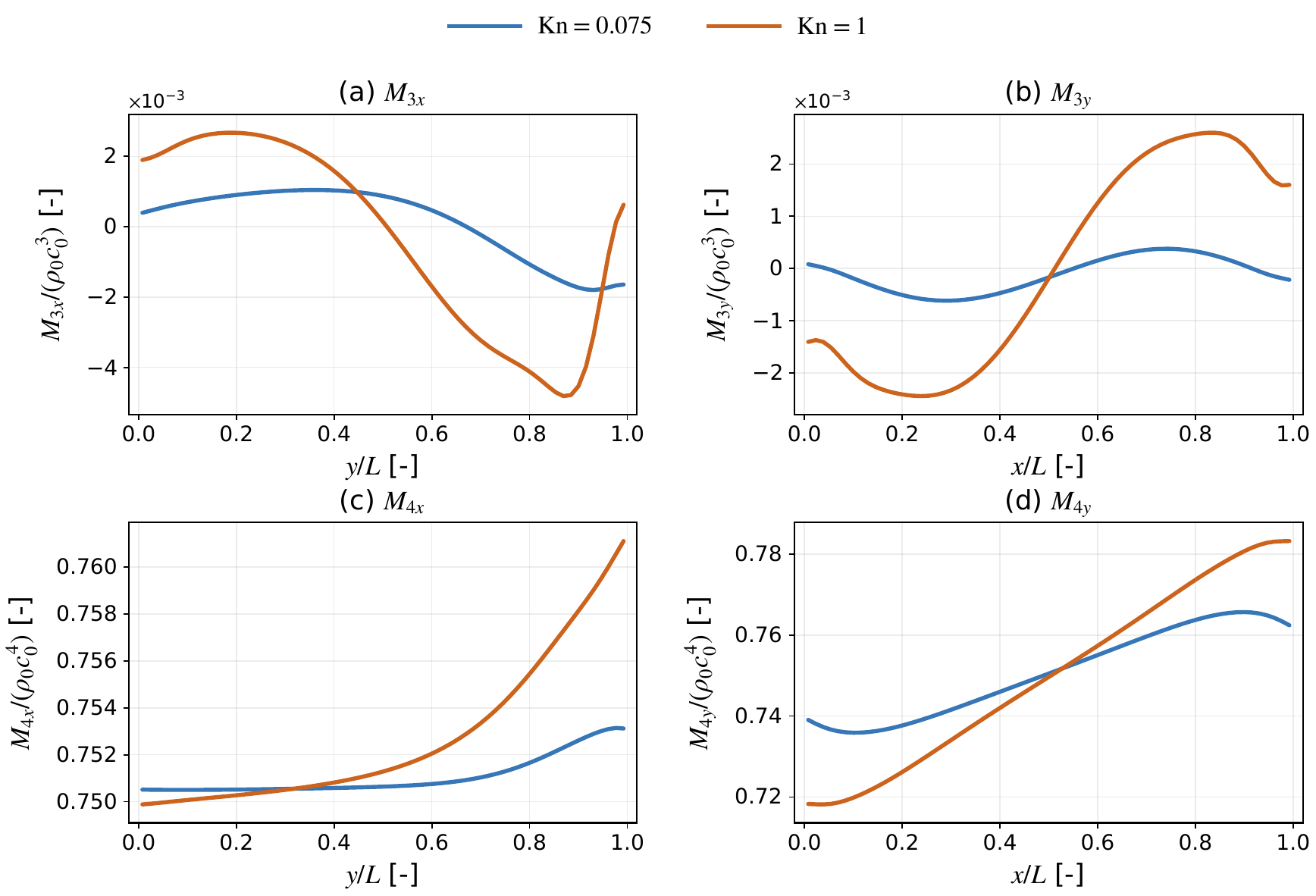}
\caption{DVM centerline profiles of the third- and fourth-order
moments at $\mathrm{Kn}=0.075$ and $1$. The streamwise moments
$M_{3x}$ and $M_{4x}$ are evaluated along the vertical centerline,
whereas $M_{3y}$ and $M_{4y}$ are evaluated along the horizontal
centerline. Increasing rarefaction substantially amplifies and
redistributes the third-order moments, revealing stronger directional
asymmetry of the molecular velocity distribution. The raw
fourth-order moments retain a large near-equilibrium contribution and
therefore vary more smoothly, while still exhibiting systematic
rarefaction-induced changes across the cavity.}
\label{fig:cavity_dvm_high_moments}
\end{figure}

The velocity, temperature, heat-flux, and shear-stress profiles in
Fig.~\ref{fig:dvm_centerlines} show that increasing rarefaction
changes both the macroscopic circulation and the associated transport
processes. A more discriminating description is obtained from the
third- and fourth-order moments, which probe directional asymmetry and
the high-energy content of the molecular velocity distribution. These
quantities are shown in Fig.~\ref{fig:cavity_dvm_high_moments} using
the converged DVM reference solutions only.

The third-order central moments measure directional asymmetry of the
molecular velocity distribution and vanish at local equilibrium. In
Fig.~\ref{fig:cavity_dvm_high_moments}, both $M_{3x}$ and $M_{3y}$
have larger amplitudes at $\mathrm{Kn}=1$ than at
$\mathrm{Kn}=0.075$. The plotted $M_{3x}$ profile changes from a
positive lower-cavity branch to a negative upper-cavity branch, while
$M_{3y}$ changes from negative on the left to positive on the right.
The zero crossings mark changes in the sign of the corresponding third
central moment. Their locations and amplitudes differ between the two
Knudsen numbers, so the higher-rarefaction profiles are not obtained by
uniformly scaling the lower-rarefaction profiles.

In Fig.~\ref{fig:cavity_dvm_high_moments}, the fourth-order moments describe the width and high-energy-tail
content of the directional molecular distributions. Unlike the
third-order moments, the raw quantities $M_{4x}$ and $M_{4y}$ contain
a substantial equilibrium background and therefore remain positive
and comparatively smooth. Nevertheless, their spatial variations
retain clear information about rarefaction. The vertical profile
$M_{4x}$ remains nearly uniform through much of the lower cavity but
increases toward the moving lid, with a considerably stronger rise at
$\mathrm{Kn}=1$. Because a fourth central moment weights large
peculiar velocities strongly, this wall-adjacent rise records a change
in the streamwise velocity spread; the profile alone is not used to
assign that change to a unique transport mechanism.

In the same figure, the transverse fourth-order moment $M_{4y}$ also changes
systematically with rarefaction. At $\mathrm{Kn}=0.075$, it exhibits a
shallow interior variation followed by an increase toward the right
side of the cavity. At $\mathrm{Kn}=1$, the profile begins at a lower
value, rises more strongly across the domain, crosses the
lower-Knudsen-number profile near the cavity center, and reaches a
larger value near the right wall. The crossing shows directly that the
change is not a uniform multiplicative increase over the plotted
centerline.

Together, Figs.~\ref{fig:dvm_centerlines} and
\ref{fig:cavity_dvm_high_moments} display the observable hierarchy used
in the subsequent error audit. The low-order fields describe the main
circulation, heat flux and stress measure transport, and the third- and
fourth-order moments add distribution-asymmetry and velocity-spread
information. The representation errors are therefore reported by
channel rather than inferred from low-order agreement alone.

\section{Gaussian representations}

\subsection{Positive phase-space model for shocks}

The shock VDF is represented in normalized phase space $\bm z=(\tilde x,\tilde\xi_x,\tilde\xi_y,\tilde\xi_z)$ by
\begin{equation}
\log \hat f_\theta(\bm z)=\log\sum_{m=1}^{N}
\exp\left[a_m-\frac12Q_m(\bm z)\right].
\label{eq:phase_gaussian}
\end{equation}
The index $m$ labels a kernel, $N$ is the number of kernels, $a_m$ is its learned log-amplitude, and $Q_m$ is a positive quadratic form. Because Eq.~\eqref{eq:phase_gaussian} is a log-sum-exp of positive Gaussian contributions, $\hat f_\theta$ is positive by construction. The diagonal form uses
\begin{equation}
Q_m(\bm z)=\sum_{\ell=1}^{4}
\left(\frac{z_\ell-\mu_{m\ell}}{s_{m\ell}}\right)^2,
\label{eq:diag_q}
\end{equation}
where $\mu_{m\ell}$ and $s_{m\ell}>0$ are the center and width of kernel $m$ in coordinate $\ell$. Equation~\eqref{eq:diag_q} restricts each kernel to axis-aligned localization in the four normalized phase coordinates. Each diagonal kernel has one amplitude, four centers, and four widths, so the parameter count is $9N$. For $N=512$, the model has 4608 parameters.

A correlated block between physical position and streamwise velocity was tested through
\begin{equation}
\begin{aligned}
Q_m^{x\xi_x}&=\frac{\alpha_m^2-2r_m\alpha_m\beta_m+\beta_m^2}{1-r_m^2},\\
\alpha_m&=\frac{\tilde x-\mu_{mx}}{s_{mx}},\qquad
\beta_m=\frac{\tilde\xi_x-\mu_{m\xi_x}}{s_{m\xi_x}}.
\end{aligned}
\label{eq:xvx}
\end{equation}
where $r_m\in(-1,1)$ is a learned correlation coefficient. Equation~\eqref{eq:xvx} adds exactly one off-diagonal degree of freedom, allowing a kernel to align with a tilted ridge in $(x,\xi_x)$ while the transverse $\xi_y$ and $\xi_z$ terms remain diagonal. This model has $10N$ parameters. Across the three tested seeds, however, it provides no clear performance advantage over the diagonal form at $N=512$ once the objective is fixed.

The main objective is a Huber loss in log-density,
\begin{equation}
\mathcal L_f=\frac{1}{B}\sum_{b=1}^{B}
H_\delta\!\left(\log\hat f_\theta(\bm z_b)-\log[f(\bm z_b)+\epsilon]\right),
\label{eq:logloss}
\end{equation}
where $B$ is the minibatch size in sampled phase points, $H_\delta$ is the Huber penalty with threshold $\delta$, and $\epsilon$ is the log floor. The pointwise log error is clipped at 20 during optimization. By operating on log density, Eq.~\eqref{eq:logloss} gives low-probability regions of the velocity distribution function (VDF) a meaningful influence without allowing numerical zeros to dominate; the Huber form also limits the leverage of isolated large residuals. For comparison, a sampled moment-informed objective adds normalized errors in selected moments at sampled spatial and velocity points. All headline moments, however, are recomputed after training by Eqs.~\eqref{eq:macro1}--\eqref{eq:m400}; no reported shock moment is taken directly from a training target.

\subsection{Mach-conditioned phase-space model}

For the nine-Mach audit, the correlated phase-space mixture is conditioned on a normalized Mach coordinate $\widetilde M$. Every unconstrained kernel parameter is represented by a degree-$d$ polynomial,
\begin{equation}
\begin{aligned}
\bm p_m(\widetilde M)&=\sum_{k=0}^{d}\bm p_{mk}\widetilde M^k,\\
\widetilde M&=2\frac{M-M_{\min}}{M_{\max}-M_{\min}}-1,
\end{aligned}
\label{eq:mach_conditioning}
\end{equation}
before the bounded center, width, and correlation transforms are applied. Equation~\eqref{eq:mach_conditioning} therefore gives every kernel parameter one shared global polynomial dependence on normalized Mach number while retaining positivity after the bounded transforms. Conditioning all ten parameters of each $x$--$\xi_x$ kernel gives $10N(d+1)$ coefficients: 15,360 for $d=2$ and 20,480 for $d=3$ at $N=512$. In float32 these shared models occupy 61,440 and 81,920 bytes. Depending on the excluded case, one model replaces 5.81--9.90 GB of eight-state phase-space data, a nominal training-database compression of approximately $7.1\times10^4$--$1.61\times10^5$.

The coefficient map is learned jointly across Mach number rather than constructed by interpolating separately fitted mixtures. That distinction is essential: Gaussian-mixture likelihoods are invariant to permutation of the kernel labels, so kernel $m$ in one independent fit has no guaranteed correspondence with kernel $m$ in another. Direct interpolation between such fits is therefore not a well-defined physical or statistical operation. A shared conditional model fixes one kernel identity across the parameter database; local or global basis functions can then move that common set of kernels without introducing label-switching ambiguity.

The targeted M6 audit replaces the global polynomial by a continuous local hat basis while retaining the same shared kernel identities,
\begin{equation}
\bm p_m(M)=\sum_{j=1}^{K}h_j(M)\bm p_{mj},
\qquad \sum_{j=1}^{K}h_j(M)=1,
\label{eq:local_mach_map}
\end{equation}
where $h_j$ is piecewise linear between ordered Mach anchors and has compact two-interval support. The three-knot map uses anchors $(1.5,5,12)$ and has $3\times10N=15{,}360$ coefficients, exactly matching the degree-2 budget. The four-knot map uses $(1.5,4,8,12)$ and has $4\times10N=20{,}480$ coefficients, exactly matching degree 3. M6 is absent from training and is decoded locally between the Mach-4 (M4) and Mach-8 (M8) anchors. Because the kernels are trained jointly, Eq.~\eqref{eq:local_mach_map} preserves component correspondence; it is not interpolation between independently permuted mixtures. This matched-budget construction isolates the effect of global versus local Mach dependence without adding representation storage.

The fitted Mach interval for either basis is computed strictly from the training cases. The conditional representation therefore carries the deterministic range guard
\begin{equation}
\mathcal A(M)=\mathbf 1\!\left(|\widetilde M|\le1\right),
\label{eq:domain_guard}
\end{equation}
where $\mathcal A=1$ authorizes decoding and $\mathcal A=0$ returns an out-of-domain flag. The M3 and M6 holdouts satisfy Eq.~\eqref{eq:domain_guard}. When M12 is excluded, $M_{\max}=8$ and $\widetilde M=2.2308$, so the query is rejected before deployment. The local basis is likewise disabled outside its terminal anchors rather than extended. The raw global-polynomial M12 evaluation is retained and reported with the conditional results as a deliberate stress test of bypassing the guard; it is not counted as an admissible model prediction.

\subsection{Normalized Gaussian moment map for cavities}

The cavity realization retains the localized support and continuous evaluation of the shock model, but acts directly on the available physical-space fields. It is
\begin{equation}
\begin{aligned}
\hat{\bm U}(\bm r)&=\bm b+\sum_{m=1}^{N}W_m(\bm r)\bm A_m,\\
G_m(\bm r)&=\exp[-(\bm r-\bm\mu_m)^T\bm D_m^{-1}(\bm r-\bm\mu_m)],\\
W_m(\bm r)&=\frac{G_m(\bm r)}{\sum_{k=1}^{N}G_k(\bm r)}.
\end{aligned}
\label{eq:cavity_map}
\end{equation}
where $\bm r=(x,y)$, $\bm b\in\mathbb R^{20}$ is a bias, $\bm A_m\in\mathbb R^{20}$ contains the amplitudes of kernel $m$ for the 20 channels, $\bm\mu_m\in\mathbb R^2$ is its center, and $\bm D_m=\mathrm{diag}(s_{mx}^2,s_{my}^2)$ contains squared widths. Equation~\eqref{eq:cavity_map} normalizes the Gaussian weights into a partition of unity shared by all 20 outputs. Each kernel has 24 parameters: two center coordinates, two widths, and 20 amplitudes. The total parameter count is therefore
\begin{equation}
P_{\mathrm{cav}}=24N+20.
\label{eq:cavity_params}
\end{equation}
Equation~\eqref{eq:cavity_params} makes the storage cost grow linearly with kernel count and includes the 20 shared output biases explicitly. The bias accounts for the field mean and each output is standardized during training. The loss is the mean squared error of the 20 normalized channels. This representation is continuous in $(x,y)$ and shares one support geometry across all 20 outputs. Its signed amplitudes match the supplied moment fields, whereas the exponential amplitudes of the phase-space realization enforce positivity of the shock VDF.

\subsection{Optimization settings}

Shock coordinates are affinely normalized by their midpoint and half-width to approximately $[-1,1]^4$. Kernel centers are parameterized as $1.15\tanh(\cdot)$, and logarithmic widths are bounded to $[-5.6,-1.25]$ through a sigmoid map. Widths are initialized at log-scale $-2.65$ and log-amplitudes at $-7.0$ with small Gaussian perturbations. Centers are initialized from $5\times10^5$ representative phase points drawn with the same mass-aware sampler used for training.

Each optimization step samples 24 physical locations and 384 velocity nodes per location. The M3 sampler uses mass exponent 0.55 and a 15\% uniform-velocity component; the M5 sampler uses exponent 0.45 and a 25\% uniform component to increase tail exposure. The Huber threshold is $\delta=1$, the pointwise log-error clip is 20, and the log floor is $10^{-35}$. Sample weights are proportional to the square root of local sampled density and clipped to $[0.25,4]$. Models are trained for 80,000 AdamW steps with default $\beta_1=0.9$, $\beta_2=0.999$, zero weight decay, gradient-norm clipping at 1, and a base learning rate of $5\times10^{-4}$. Center and width learning rates are multiplied by 0.5 and 0.25, respectively; checkpoints are written every 2000 steps. Table~\ref{tab:training} summarizes these settings. Shock training was performed on NVIDIA Tesla M40 24-GB GPUs.

The conditional models use the same $24\times384$ phase-point minibatch, mass exponent 0.55, 15\% uniform-velocity sampling, and $10^{-35}$ floor. At each step one of the eight training Mach cases is selected, its physical locations and velocities are sampled, the selected global or local basis generates the kernel parameters, and the log-density loss is evaluated. Every tenth step, a moment penalty of weight 0.01 is computed from 12 physical locations and 1536 velocity samples for $\rho$, $u_x$, $T$, $q_x$, $\sigma_{xx}$, and $M_{300}^{\mathrm{neq}}$. The 18 global-polynomial models and six local-M6 models are each trained for 160,000 steps with a 1000-step warmup, base learning rate $5\times10^{-4}$, terminal learning-rate ratio 0.05, and gradient clipping at 1. Seeds 1234, 2026, and 3407 are used for every comparison. Table~\ref{tab:conditional_protocol} records the held-out cases, bases, budgets, and protocols. Evaluation uses 50,000 sampled phase points and full spatial moment profiles; all six local runs finished normally and use the same terminal selection rule as the global audit.

The complete training scheme is therefore: lock and checksum the DVM states; normalize coordinates using training cases only; initialize centers from the mass-aware pool; alternate stochastic log-density steps with the prescribed moment audit; save fixed-interval checkpoints; select the terminal protocol without inspecting holdout errors; and finally reconstruct all moments with the reference quadrature. This sequence, together with all numerical hyperparameters, is stored in the supplied JavaScript Object Notation (JSON) configurations and histories.

Cavity outputs are standardized channel by channel before optimization and transformed back to physical units for all reported errors. The cavity models use batch size 4096, 80,000--100,000 Adam steps, gradient clipping at 1, and learning rates from $7\times10^{-4}$ to $10^{-3}$. Table~\ref{tab:training} records the principal settings, and the accompanying archive contains the machine-readable configuration files.

\begin{table}[t]
\caption{Principal optimization settings. For shocks, ``uniform fraction'' denotes the fraction of velocity samples drawn uniformly rather than from the mass-weighted distribution.}
\label{tab:training}
\centering
\small
\begin{tabular*}{\textwidth}{@{\extracolsep{\fill}}lcccccc@{}}
\toprule
Shock model & $N$ & steps & base rate & $x\times\xi$ samples/step & center pool & uniform fraction\\
\midrule
M3 diagonal & 512 & 80,000 & $5\times10^{-4}$ & $24\times384$ & $5\times10^5$ & 0.15\\
M5 diagonal & 512 & 80,000 & $5\times10^{-4}$ & $24\times384$ & $5\times10^5$ & 0.25\\
M5 $x$--$\xi_x$ & 512 & 80,000 & $5\times10^{-4}$ & $24\times384$ & $5\times10^5$ & 0.25\\
\bottomrule
\end{tabular*}
\vspace{5pt}

\begin{tabular*}{0.82\textwidth}{@{\extracolsep{\fill}}lccccc@{}}
\toprule
Cavity case & $N$ & steps & learning rate & batch size & center init.\\
\midrule
$\mathrm{Kn}=0.075$ & 256/512 & 80,000 & $0.9$--$1.0\times10^{-3}$ & 4096 & random\\
$\mathrm{Kn}=1$ & 256/512 & 100,000 & $0.7$--$0.8\times10^{-3}$ & 4096 & random\\
\bottomrule
\end{tabular*}
\end{table}

\begin{table}[t]
\caption{Conditional Mach protocol. Each row uses three seeds, producing 18 global-polynomial runs and six matched-budget local-M6 runs.}
\label{tab:conditional_protocol}
\centering
\small
\begin{tabular}{lccccc}
\toprule
Holdout & role & train range & degree & parameters & steps\\
\midrule
M3 & interpolation & $[1.5,12]$ & 2 / 3 & 15,360 / 20,480 & 160,000\\
M6 & interpolation & $[1.5,12]$ & 2 / 3 & 15,360 / 20,480 & 160,000\\
M12 & out-of-domain audit & $[1.5,8]$ & 2 / 3 & 15,360 / 20,480 & 160,000\\
M6 local & interpolation & $[1.5,12]$ & 3 / 4 knots & 15,360 / 20,480 & 160,000\\
\bottomrule
\end{tabular}
\end{table}

Table~\ref{tab:conditional_protocol} separates three distinct tests under identical three-seed and 160,000-step protocols. M3 and M6 are bracketed interpolation holdouts, M12 is an explicitly out-of-domain audit because removing it truncates the training interval at M8, and the local-M6 rows preserve the two global parameter budgets while changing only the Mach-coordinate basis. This organization makes basis locality, rather than extra storage or additional optimization, the controlled variable in the M6 comparison.

\section{Evaluation and comparison protocols}

\subsection{Error norms and storage accounting}

For a reference profile or field $g$ and reconstruction $\hat g$, the principal metric is
\begin{equation}
E_g=\frac{\|\hat g-g\|_2}{\|g\|_2}.
\label{eq:error}
\end{equation}
For shock profiles, the norm is taken over physical stations after the VDF has been integrated by the DVM quadrature. For cavity fields, the norm is over the $65\times65$ physical grid. Contour error maps use the signed range-normalized percentage
\begin{equation}
e_g(\bm r)=100\frac{\hat g(\bm r)-g(\bm r)}{\max g-\min g},
\label{eq:rangeerror}
\end{equation}
Equation~\eqref{eq:rangeerror} remains finite for fields that cross zero and preserves the sign of local over- or under-prediction; it is therefore used only for the spatial error maps. The maximum cavity error quoted in summary tables is instead the largest global $E_g$ among all 20 channels.

Nominal compression is defined as the number of stored reference scalar values divided by the number of representation parameters,
\begin{equation}
C_{\mathrm{nom}}=\frac{N_{\mathrm{raw}}}{P}.
\label{eq:compression}
\end{equation}
Equation~\eqref{eq:compression} places grid and Gaussian representations on the same coefficient-count basis. This measure is transparent but intentionally nominal: a production archive may exploit lossless compression, symmetry, or reduced velocity coordinates. The comparison here asks a simpler and reproducible question: with the same number of floating-point coefficients, which representation preserves the kinetic observables most accurately?

\subsection{Seed and checkpoint protocol}

Each M5 architecture was trained from three independent pseudorandom
initializations. For a paired comparison, the same random seeds
(1234, 2026, and 3407) were used for the diagonal and correlated
models. The seeds control Gaussian-center initialization, stochastic
phase-space sampling, and minibatch construction. Results are reported from the terminal checkpoint at 80,000 steps for every seed. Table~\ref{tab:checkpoint} compares the terminal checkpoint with the checkpoint having the smallest recorded training loss. The audit shows that the lowest noisy minibatch loss is not consistently the lowest moment error, especially for $q_x$. A fixed terminal rule is therefore applied uniformly to all architecture statistics, keeping model selection independent of the reported observables.

\subsection{Held-out spatial interpolation}

To test whether the continuous representation merely memorizes stored shock stations, every fifth interior M5 station is excluded from all training sampling and from the center-initialization pool. This leaves 1760 training and 440 held-out stations. Errors are evaluated separately on the two sets, both globally and in the steep region $|x/\lambda|\le10$. The test directly measures interpolation within the fixed M5 state and isolates continuity with respect to the original storage grid.

\subsection{Leave-one-Mach-out protocol}

The parametric test is stricter than the station holdout because an entire kinetic state is removed. For each holdout in $\{\mathrm{M3},\mathrm{M6},\mathrm{M12}\}$, separate degree-2 and degree-3 models are trained from three seeds on the other eight Mach cases. A subsequent matched-budget basis audit repeats the M6 exclusion with three- and four-knot local maps over the same three seeds. The reported distribution metric is a relative $L_2$ error over 50,000 phase points sampled by the common evaluation manifest; the weighted log-density root-mean-square error is recorded separately. Moment errors use Eq.~\eqref{eq:error} over the complete stored spatial profile. Training-case statistics include only cases actually seen by a given model. Holdout values are never used for checkpoint selection, coordinate normalization, or terminal-model selection. M12 queries are not clipped to the training interval, so loss of physical validity is visible in the reported profiles.

\subsection{Matched-storage shock grids}

The shock baseline stores $\log f$ on coarse four-dimensional grids in $(x,\xi_x,\xi_y,\xi_z)$ and reconstructs it by multilinear interpolation before applying the same DVM moment quadrature. Table~\ref{tab:grid_layouts} gives the tested layouts at two budgets: 4608 values, matching the diagonal $N=512$ model, and 5120 values, matching the correlated model. The layouts are a balanced uniform grid; a physics-oriented uniform grid allocating more nodes to $x$ and $\xi_x$; and a physics-oriented grid with adaptive $x$ locations. The adaptive baseline is intentionally generous: the $x$ points are placed using gradients and transport information from the reference. It is therefore an oracle, physics-informed regular-grid baseline.

\subsection{Cavity bilinear and truncated-SVD baselines}

Two conventional reconstructions are used as matched-storage baselines for the cavity. The first is bilinear interpolation on a coarse uniform grid. Each of the 20 moment fields is sampled on the same coarse grid, whose resolution is selected so that the total number of stored field values remains slightly below the number of parameters in the corresponding Gaussian model.

The second baseline uses a separate truncated singular-value decomposition (SVD) for each normalized $65\times65$ field. All 20 fields are reconstructed with the same retained rank. For comparison with the 6164-parameter Gaussian model, rank 2 is used, requiring 5260 stored coefficients in total. For the 12,308-parameter Gaussian model, rank 4 is used, requiring 10,500 coefficients. Because the decomposition is applied independently to each steady field, this baseline is more accurately described as a per-field truncated SVD than as snapshot-based proper orthogonal decomposition (POD).\cite{Sirovich1987,Berkooz1993,TrefethenBau1997}

\section{Shock-wave results}

\subsection{Mach-3 reconstruction and the role of the objective}

\begin{figure}[!t]
\centering
\includegraphics[width=\textwidth]{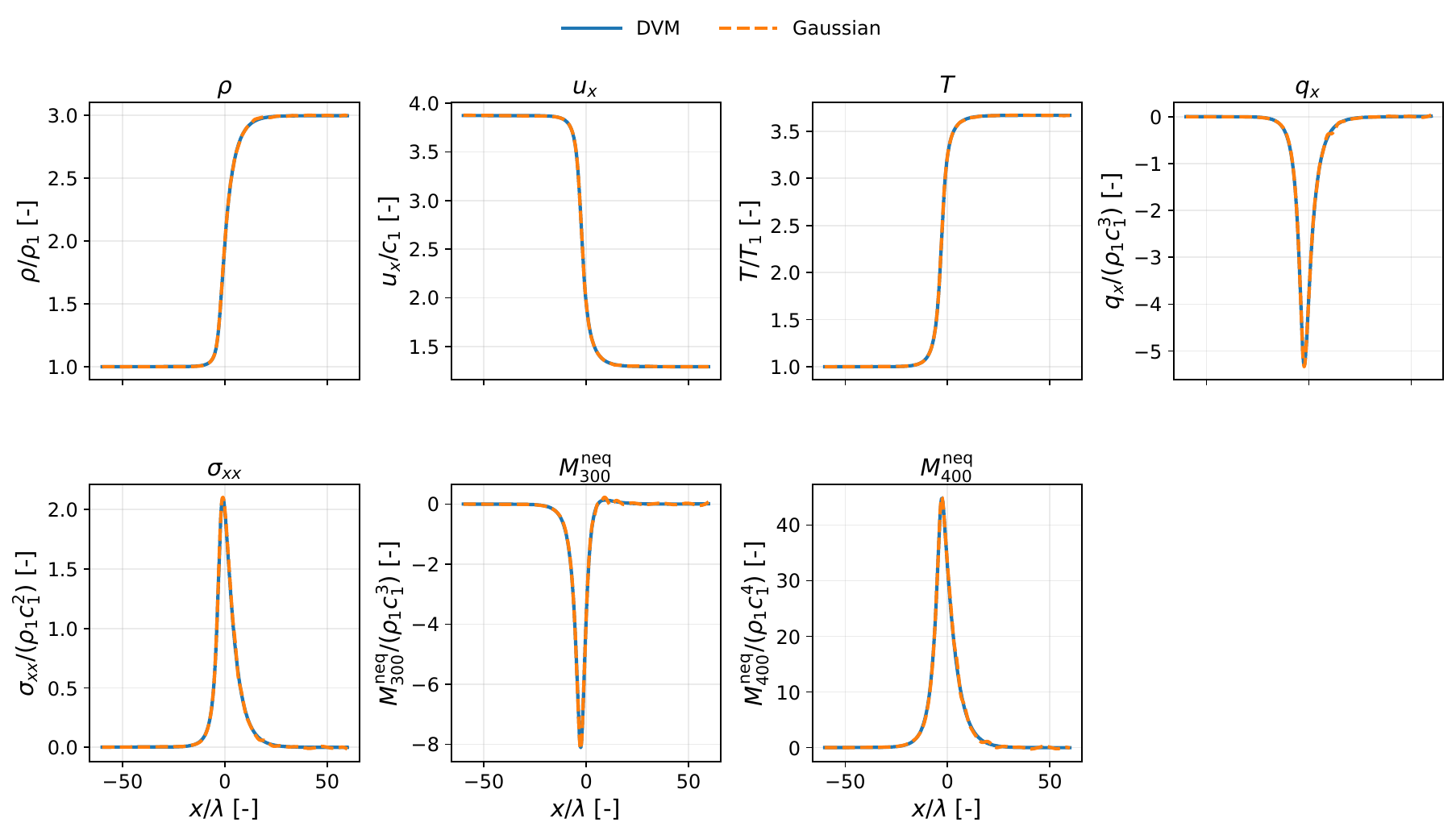}
\caption{Mach-3 normal shock reconstructed by the diagonal $N=512$ positive phase-space model trained on $\log f$. The seven panels show low-order fields, transport moments, and third- and fourth-order nonequilibrium deviations.}
\label{fig:m3}
\end{figure}

Figure~\ref{fig:m3} compares the M3 DVM profiles with a diagonal $N=512$ Gaussian model trained only on $\log f$.

\FloatBarrier

The low-order jump conditions are nearly indistinguishable: $E_\rho=0.142\%$, $E_{u_x}=0.0656\%$, and $E_T=0.0648\%$. More importantly, the localized heat-flux minimum, normal-stress maximum, negative third moment, and fourth-order deviation are preserved with errors of 1.48\%, 1.68\%, 2.03\%, and 2.07\%, respectively. The fourth-order curve is especially informative because the equilibrium contribution has been subtracted; the agreement is not produced by a large common background.

The controlled M3 ablation in Table~\ref{tab:m3_ablation} shows a larger change from the tested objective than from the tested covariance or capacity modifications. With the same diagonal $N=512$ model, sampled moment-informed training gives stress and heat-flux errors of 17.84\% and 7.97\%, compared with 1.68\% and 1.48\% for log-density training. Adding an $x$--$\xi_x$ block under the sampled-moment objective does not reduce those two errors, and the tested $N=1024$ variant increases $q_x$ and $M_{300}^{\mathrm{neq}}$ errors. These data support log-density training for the present M3 configuration. They do not separately identify whether the difference arises from phase-space coverage, stochastic moment quadrature, or optimization, so no universal ranking of objectives is inferred.

\begin{table}[t]
\caption{M3 relative errors in percent for the objective and capacity ablation. The log-density row uses the terminal checkpoint. A dash indicates that the fourth-order deviation was not stored for the earlier moment-loss run.}
\label{tab:m3_ablation}
\centering
\begin{tabular}{lccccccc}
\toprule
Configuration & $E_\rho$ & $E_{u_x}$ & $E_T$ & $E_{q_x}$ & $E_{\sigma_{xx}}$ & $E_{M_{300}}$ & $E_{M_{400}^{\rm neq}}$\\
\midrule
Diagonal 512, $\log f$ & 0.142 & 0.0656 & 0.0648 & 1.475 & 1.681 & 2.030 & 2.066\\
Diagonal 512, moment loss & 0.935 & 0.306 & 0.514 & 7.969 & 17.843 & 8.235 & --\\
$x$--$\xi_x$ 512, moment loss & 0.934 & 0.300 & 0.476 & 8.058 & 17.266 & 8.644 & --\\
$x$--$\xi_x$ 1024, moment loss & 0.940 & 0.280 & 0.541 & 10.423 & 17.762 & 10.515 & --\\
\bottomrule
\end{tabular}
\end{table}

\subsection{Mach-5 reconstruction}

\begin{figure}[!tbp]
\centering
\includegraphics[width=\textwidth]{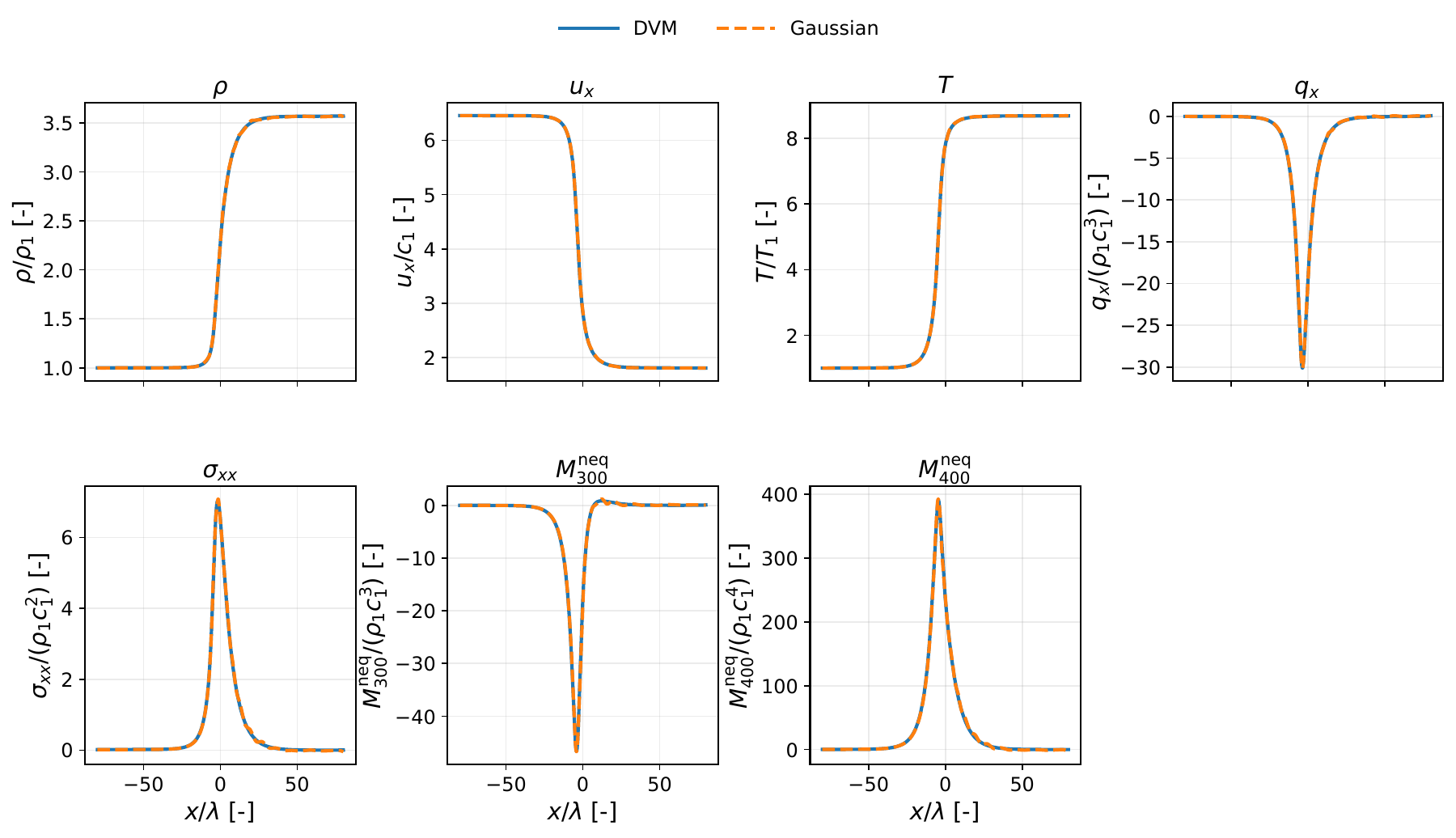}
\caption{Mach-5 normal shock reconstructed by the diagonal $N=512$ log-density model. DVM and reconstructed profiles are shown for the low-order, transport, and higher-moment observables used in the error calculation.}
\label{fig:m5}
\end{figure}

The stronger M5 shock provides a larger dynamic-range test. Figure~\ref{fig:m5} uses the diagonal $N=512$ model and seed 1234. The representative errors are $0.231\%$, $0.0752\%$, and $0.0900\%$ for $\rho$, $u_x$, and $T$, and 1.12\%, 1.66\%, 1.71\%, and 1.80\% for $q_x$, $\sigma_{xx}$, $M_{300}^{\mathrm{neq}}$, and $M_{400}^{\mathrm{neq}}$. The plotted fourth-order peak is roughly two orders of magnitude larger than in the M3 case, and the error remains 1.80\% for this observable.

For M5, Fig.~\ref{fig:m5} supports the reported accuracy of the selected log-density run. The controlled objective comparison is confined to M3 in Table~\ref{tab:m3_ablation}; consequently, no M5 objective-improvement factor is inferred here.

\subsection{Architecture, seeds, and checkpoint behavior}

\begin{figure}[!tbp]
\centering
\includegraphics[width=\textwidth]{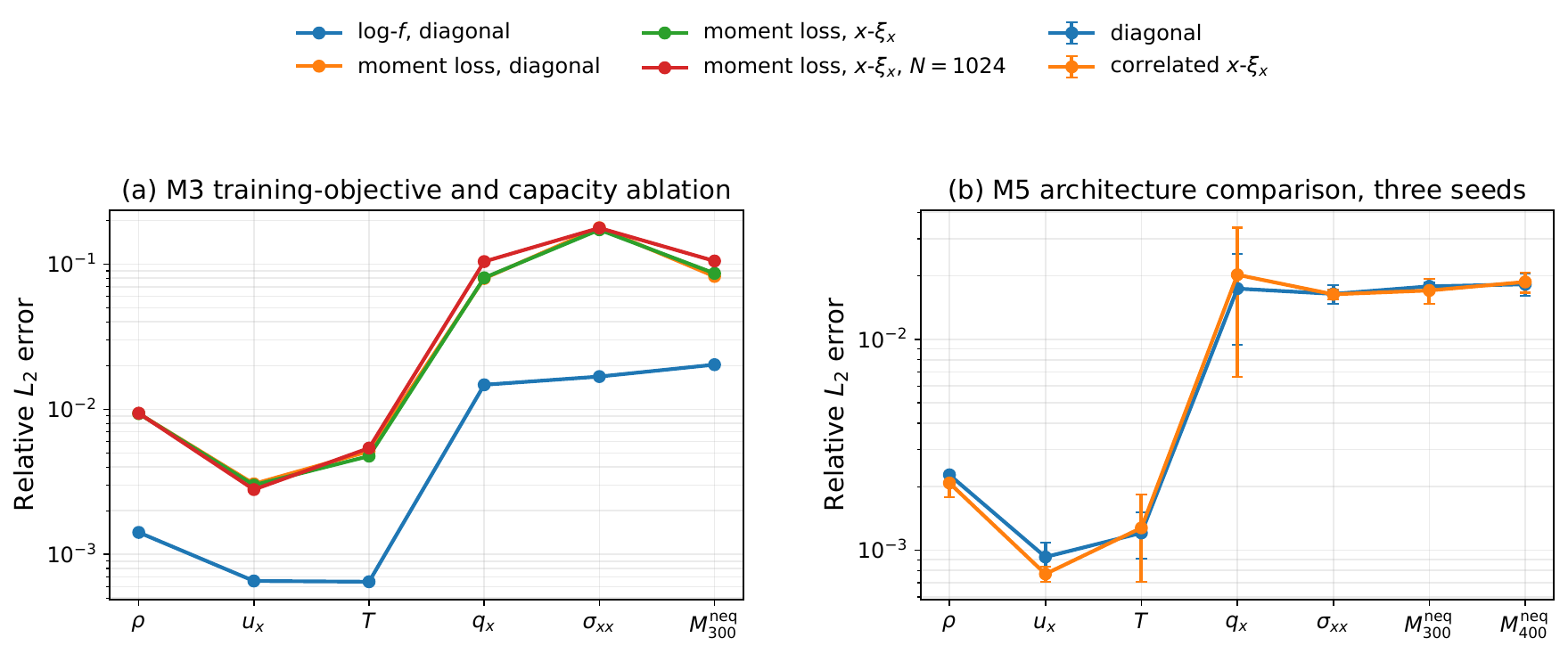}
\caption{Ablation results. (a) For M3, changing the objective from sampled moment supervision to direct log-density fitting reduces the displayed transport and higher-moment errors by factors of approximately 4--11. The tested $N=1024$ moment-loss run does not reduce these errors. (b) For M5 at $N=512$, the diagonal and correlated $x$--$\xi_x$ models are compared across three seeds; error bars are one sample standard deviation.}
\label{fig:ablation}
\end{figure}

Figure~\ref{fig:ablation} places the M3 objective test and the M5 architecture comparison on the same error axes. For M5, the diagonal and $x$--$\xi_x$ models are each evaluated over three seeds. Their mean nonequilibrium errors differ by less than one sample standard deviation, as listed in Table~\ref{tab:m5_seeds}. The correlated model has a slightly smaller mean stress error, while the diagonal model has slightly smaller mean $q_x$ and $M_{400}^{\mathrm{neq}}$ errors. With only three seeds, these data support no formal performance ranking. The diagonal model is used for the representative profile because it has 10\% fewer parameters and comparable errors in this sample.

\begin{table}[t]
\caption{M5 terminal-checkpoint errors in percent over seeds 1234, 2026, and 3407. Values are mean $\pm$ sample standard deviation. Splitting the low-order and nonequilibrium blocks keeps the numerical precision readable.}
\label{tab:m5_seeds}
\centering
\small
\begin{tabular}{lccc}
\toprule
Architecture & $E_\rho$ & $E_{u_x}$ & $E_T$\\
\midrule
Diagonal & $0.2275\pm0.0077$ & $0.0927\pm0.0156$ & $0.1206\pm0.0299$\\
$x$--$\xi_x$ & $0.2081\pm0.0293$ & $0.0770\pm0.0062$ & $0.1275\pm0.0566$\\
\bottomrule
\end{tabular}
\vspace{5pt}

\begin{tabular}{lcccc}
\toprule
Architecture & $E_{q_x}$ & $E_{\sigma_{xx}}$ & $E_{M_{300}^{\rm neq}}$ & $E_{M_{400}^{\rm neq}}$\\
\midrule
Diagonal & $1.740\pm0.796$ & $1.642\pm0.165$ & $1.784\pm0.085$ & $1.823\pm0.221$\\
$x$--$\xi_x$ & $2.023\pm1.361$ & $1.634\pm0.090$ & $1.704\pm0.236$ & $1.871\pm0.208$\\
\bottomrule
\end{tabular}
\end{table}

\begin{figure}[!tbp]
\centering
\includegraphics[width=\textwidth]{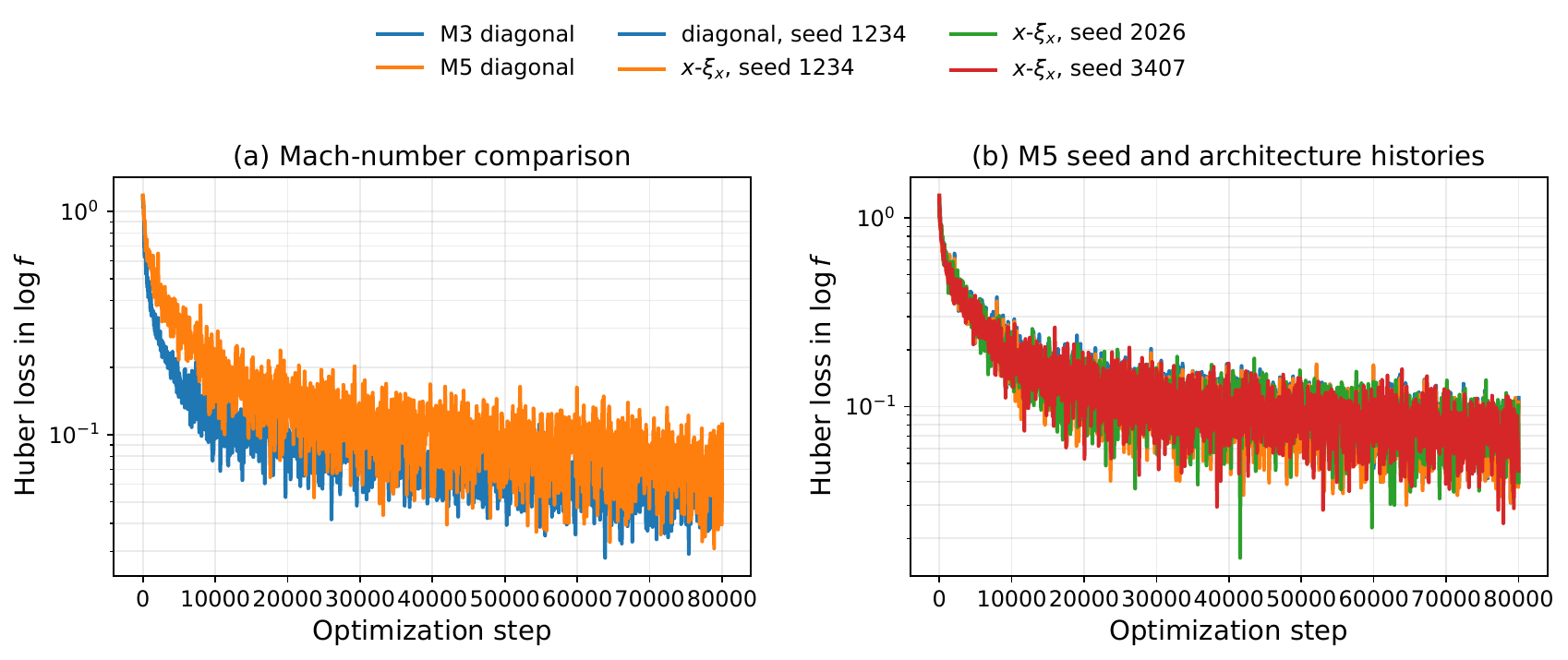}
\caption{Shock optimization histories. (a) M3 and M5 diagonal models. (b) M5 diagonal and correlated runs with different seeds. The minibatch Huber loss remains noisy after the rapid initial decay; terminal checkpoints are used uniformly rather than selecting a checkpoint by a reported moment.}
\label{fig:training}
\end{figure}

The heat-flux error has the largest sample standard deviation in Table~\ref{tab:m5_seeds}. One seed gives $E_{q_x}=3.59\%$ for the correlated model and 2.64\% for the diagonal model, whereas the other seeds are close to 1.2--1.7\%. The variation is not consistently removed by choosing the checkpoint with the smallest stored minibatch loss; for some runs that choice makes $q_x$ worse. Figure~\ref{fig:training} shows noisy late-stage loss histories, but those histories do not identify a unique cause for the seed variation. A fixed terminal checkpoint is therefore used for every seed rather than selecting by a reported moment.

\subsection{Held-out spatial interpolation}

\begin{figure}[!tbp]
\centering
\includegraphics[width=\textwidth]{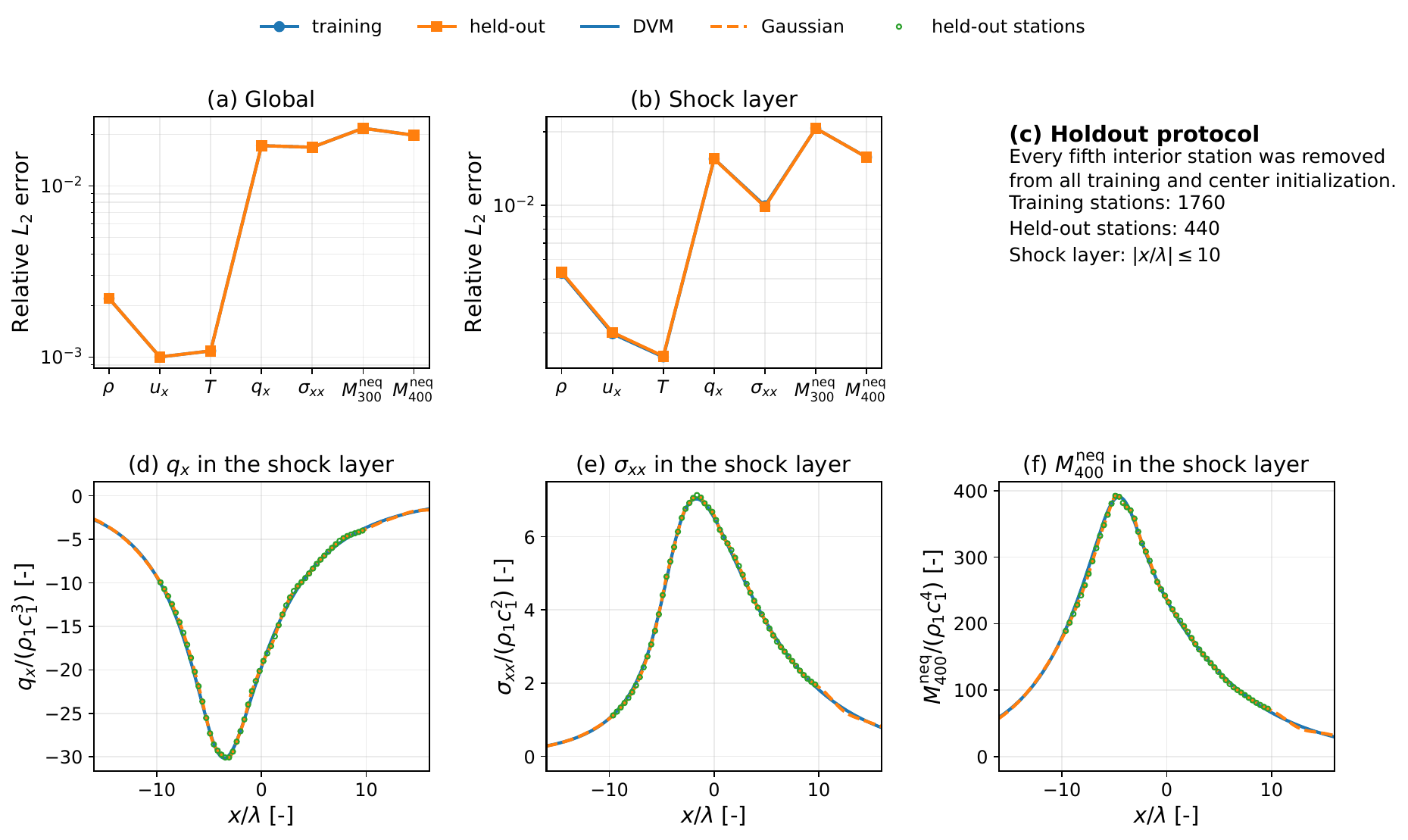}
\caption{Held-out-station interpolation for the M5 shock. Panels (a) and (b) compare relative errors on training and unseen stations globally and within the shock layer; panel (c) states the exclusion protocol. Panels (d)--(f) show sensitive profiles and mark held-out stations inside $|x/\lambda|\le10$. The unseen points retain the profile accuracy of the training set.}
\label{fig:holdout}
\end{figure}

The held-out experiment tests interpolation between stored physical stations within one fixed M5 state. Global train and held-out errors agree to at least four significant digits. In the shock layer, Table~\ref{tab:holdout} and Fig.~\ref{fig:holdout} show that the training--holdout difference is at most 1.3\% of the corresponding training error. The test therefore supports spatial interpolation on this interleaved M5 grid; it does not by itself establish accuracy for a different shock state or a different spatial sampling pattern.

\begin{table}[t]
\caption{M5 holdout errors in percent. Every fifth interior station is excluded from training and center initialization. The shock-layer norm uses $|x/\lambda|\le10$.}
\label{tab:holdout}
\centering
\begin{tabular}{llccccccc}
\toprule
Region & split & $E_\rho$ & $E_{u_x}$ & $E_T$ & $E_{q_x}$ & $E_{\sigma_{xx}}$ & $E_{M_{300}}$ & $E_{M_{400}^{\rm neq}}$\\
\midrule
Global & training & 0.2200 & 0.1002 & 0.1086 & 1.7181 & 1.6811 & 2.1706 & 1.9767\\
Global & held out & 0.2199 & 0.1002 & 0.1086 & 1.7181 & 1.6811 & 2.1706 & 1.9767\\
Shock layer & training & 0.5272 & 0.2982 & 0.2396 & 1.5507 & 0.9957 & 2.0657 & 1.5752\\
Shock layer & held out & 0.5318 & 0.3019 & 0.2413 & 1.5502 & 0.9859 & 2.0680 & 1.5726\\
\bottomrule
\end{tabular}
\end{table}

\subsection{Matched-storage regular-grid comparison}

\begin{figure}[!tbp]
\centering
\includegraphics[width=\textwidth]{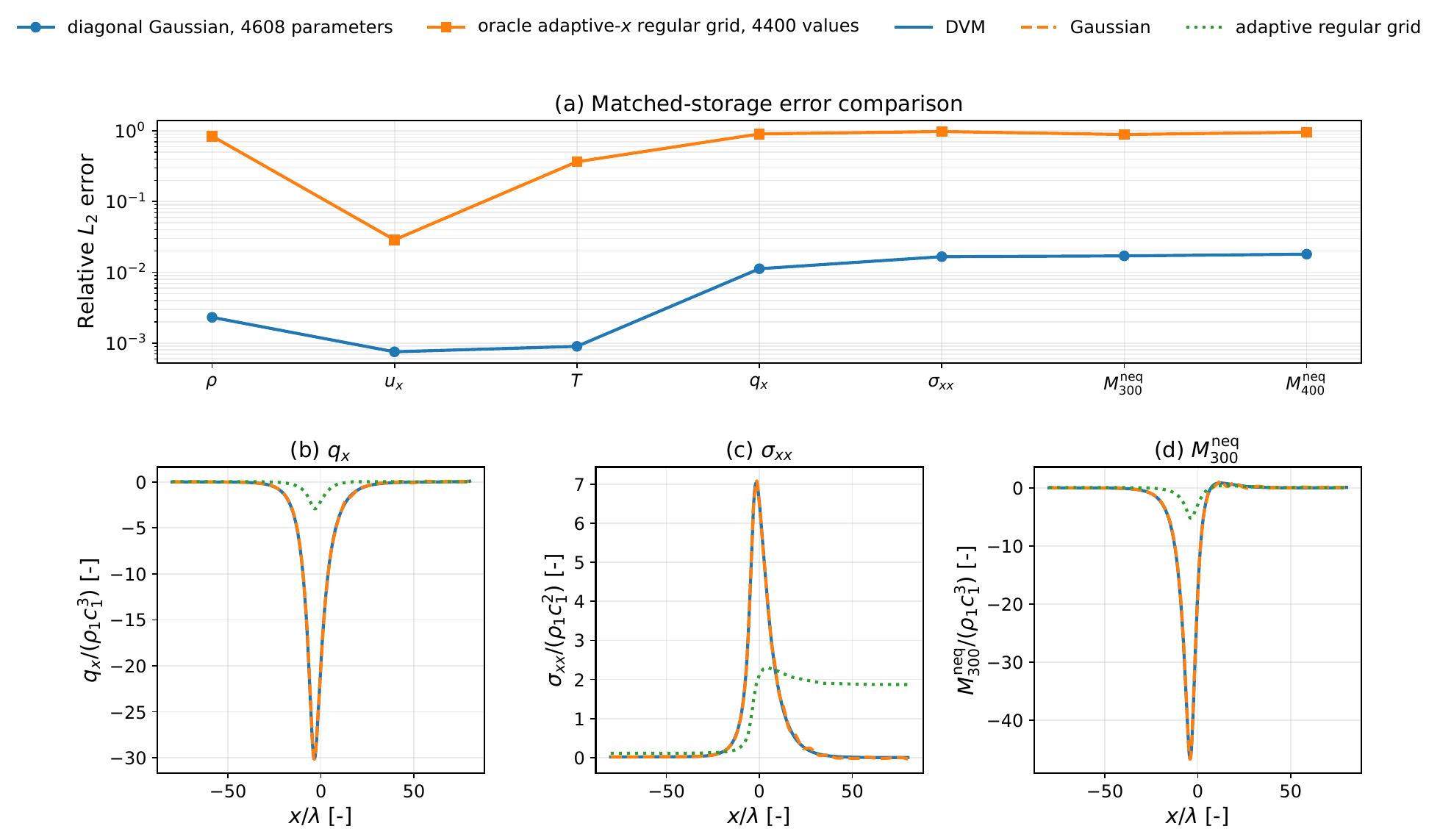}
\caption{M5 matched-storage comparison. (a) Relative errors of the diagonal Gaussian model and the best tested conventional grid, an oracle physics-informed grid with adaptive $x$ placement. (b)--(d) Heat-flux, stress, and third-order profiles. At this 4400-value grid budget, adaptive physical-node placement does not prevent large moment errors.}
\label{fig:shock_baseline}
\end{figure}

The most demanding test asks whether Gaussian localization provides value beyond storing $\log f$ on a coarse grid. At the diagonal-model budget, a four-dimensional regular grid has only 4608 values to distribute among one physical and three velocity coordinates. Even the oracle adaptive-$x$ layout stores 4400 values and uses reference gradients to concentrate spatial nodes near the shock. It still produces 84.3\% density error, 36.6\% temperature error, and 89--98\% errors in the transport and high-order moments. The corresponding Gaussian errors are 0.075--0.231\% for low-order fields and 1.12--1.80\% for nonequilibrium quantities. Figure~\ref{fig:shock_baseline} compares the profiles, Table~\ref{tab:shock_baseline} reports the headline moment set, and Table~\ref{tab:grid_layouts} documents both coefficient budgets and all three grid layouts.

\begin{table}[t]
\caption{M5 matched-storage errors in percent. The Gaussian has 4608 parameters and nominal compression $4.91\times10^4$; the oracle adaptive-$x$ regular grid stores 4400 values and has nominal compression $5.14\times10^4$.}
\label{tab:shock_baseline}
\centering
\begin{tabular}{lccccccc}
\toprule
Method & $E_\rho$ & $E_{u_x}$ & $E_T$ & $E_{q_x}$ & $E_{\sigma_{xx}}$ & $E_{M_{300}}$ & $E_{M_{400}^{\rm neq}}$\\
\midrule
Diagonal Gaussian & 0.231 & 0.075 & 0.090 & 1.125 & 1.663 & 1.708 & 1.804\\
Adaptive regular grid & 84.285 & 2.884 & 36.586 & 90.440 & 97.884 & 88.712 & 95.761\\
\bottomrule
\end{tabular}
\end{table}

\begin{table}[t]
\caption{Regular-grid layout audit for both matched-storage budgets. Errors are percentages. ``Balanced'' distributes nodes nearly uniformly; ``physics uniform'' favors $x$ and $\xi_x$; ``adaptive $x$'' places physical nodes using reference gradients and transport profiles.}
\label{tab:grid_layouts}
\centering
\small
\begin{tabular}{rrlrrrrr}
\toprule
Budget & stored & layout & $E_{q_x}$ & $E_{\sigma_{xx}}$ & $E_{M_{300}^{\rm neq}}$ & $E_{M_{400}^{\rm neq}}$ & max.\\
\midrule
4608 & 4608 & balanced & 98.776 & 97.471 & 102.515 & 97.827 & 102.515\\
4608 & 4400 & physics uniform & 89.064 & 102.957 & 87.445 & 97.126 & 102.957\\
4608 & 4400 & adaptive $x$ & 90.440 & 97.884 & 88.712 & 95.761 & 97.884\\
5120 & 5120 & balanced & 100.658 & 97.944 & 103.745 & 98.361 & 103.745\\
5120 & 5000 & physics uniform & 80.461 & 95.219 & 80.124 & 92.106 & 95.219\\
5120 & 5000 & adaptive $x$ & 79.464 & 92.792 & 79.238 & 91.093 & 92.792\\
\bottomrule
\end{tabular}
\end{table}

The layout audit establishes a consistent matched-budget advantage over all three tested multilinear regular-grid layouts. Tensor decompositions, reduced velocity coordinates, and nonlinear encoders provide complementary approximation structures for future coefficient-matched benchmarks.

\subsection{Parametric shock strength and range-guarded deployment}

\begin{figure}[!tbp]
\centering
\includegraphics[width=\textwidth]{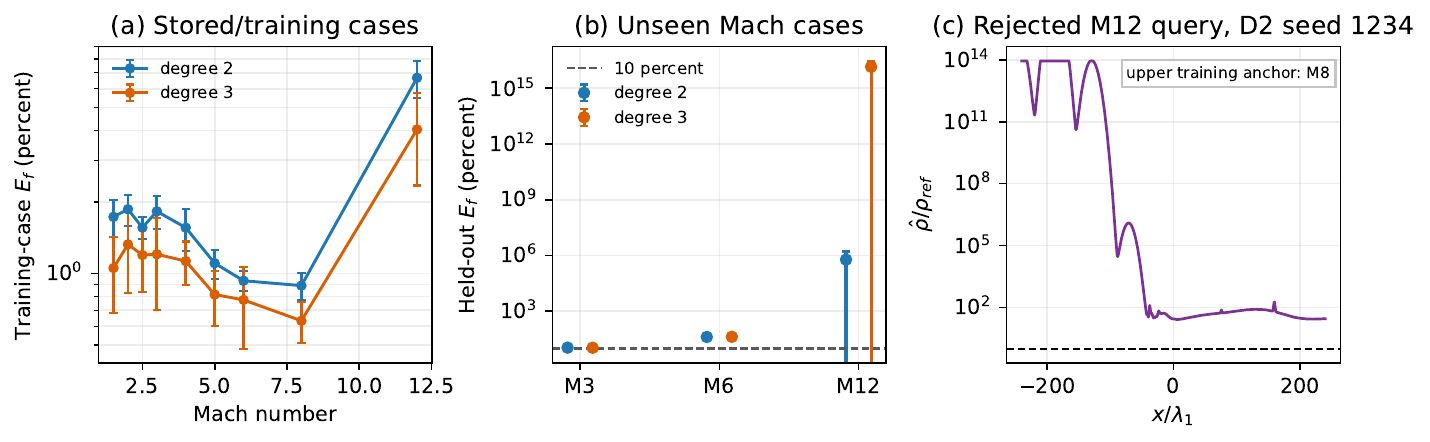}
\caption{Mach-conditioned range and holdout audit. (a) Training-case sampled distribution errors over the nine-Mach database. (b) Whole-state M3 and M6 tests measure in-range holdout errors; the M12 values are unguarded evaluations of a query rejected by Eq.~\eqref{eq:domain_guard}. (c) A representative raw degree-2 M12 density ratio shows the result of bypassing the guard beyond the M8 training boundary; the dashed line marks unity. Error bars are one sample standard deviation over three seeds. Moment-resolved in-range errors are reported in Table~\ref{tab:mach_holdout}.}
\label{fig:full_mach}
\end{figure}

Figure~\ref{fig:full_mach}(a) summarizes the 18 completed global conditional runs. Averaged over the cases used in training, the sampled distribution error is 1.87\% for degree 2 and 1.27\% for degree 3. The error is 0.89\% and 0.63\% at M8 and 6.67\% and 4.04\% at the trained M12 endpoint. Complete-state holdouts in Fig.~\ref{fig:full_mach}(b) provide the corresponding interpolation test.

Table~\ref{tab:mach_holdout} reports the bracketed M3 and M6 holdouts. At M3, both global degrees give approximately 11\% sampled distribution error, 13.9--14.7\% density error, and 23--42\% transport and higher-moment errors. At M6, the global maps give approximately 42\% sampled distribution error. The contrast with the training-state errors identifies whole-state interpolation, especially transport recovery, as the more demanding task; increasing the global degree mainly redistributes error among channels.

Table~\ref{tab:mach_holdout} also isolates the effect of the coefficient basis at M6. The four-knot map reduces the distribution error from $42.86\pm5.40\%$ for global degree 3 to $11.45\pm0.94\%$ at the identical 20,480-parameter budget, a 73.3\% reduction. Against the grid-refined fine M6 reference, errors fall from 21.42\% to 6.96\% in density, 6.85\% to 3.05\% in velocity, 7.37\% to 5.53\% in temperature, 56.54\% to 37.98\% in stress, and 46.31\% to 30.11\% in the fourth-order nonequilibrium moment. The mean heat-flux error changes from 38.30\% to 39.50\%, a 1.20-point increase smaller than the reported sample standard deviations. The K3 run has an intermediate distribution error but larger moment errors; because K3 and K4 differ in both anchor placement and coefficient count, their contrast does not isolate a single mechanism. These results identify local Mach dependence as a major source of improvement while retaining transport moments as the controlling acceptance criterion.

\begin{table}[t]
\caption{Leave-one-Mach-out errors in percent, mean $\pm$ sample standard deviation over three seeds. M6 moments are re-scored against the grid-refined fine reference. D2 and D3 denote global degree-2 and degree-3 maps; K3 and K4 denote three- and four-knot correspondence-preserving local maps.}
\label{tab:mach_holdout}
\centering
\scriptsize
\resizebox{\textwidth}{!}{%
\begin{tabular}{llcccccccc}
\toprule
Case & map & $E_f$ & $E_\rho$ & $E_{u_x}$ & $E_T$ & $E_{q_x}$ & $E_{\sigma_{xx}}$ & $E_{M_{300}}$ & $E_{M_{400}^{\rm neq}}$\\
\midrule
M3 & global D2 & $11.18\pm1.38$ & $13.86\pm0.44$ & $3.07\pm0.31$ & $8.83\pm0.09$ & $23.16\pm3.70$ & $24.34\pm4.86$ & $31.11\pm3.53$ & $29.13\pm2.20$\\
M3 & global D3 & $11.04\pm0.40$ & $14.67\pm0.05$ & $3.75\pm0.21$ & $8.59\pm0.14$ & $36.38\pm4.69$ & $30.50\pm3.85$ & $41.81\pm5.42$ & $40.71\pm2.75$\\
M6 & global D2 & $41.90\pm6.24$ & $17.50\pm1.52$ & $4.85\pm0.80$ & $5.79\pm0.11$ & $54.24\pm3.39$ & $73.61\pm9.45$ & $62.28\pm8.71$ & $72.54\pm11.87$\\
M6 & local K3 & $21.85\pm4.06$ & $24.68\pm2.98$ & $5.58\pm1.78$ & $7.50\pm1.34$ & $66.24\pm12.75$ & $70.24\pm32.91$ & $77.17\pm7.04$ & $83.10\pm11.26$\\
M6 & global D3 & $42.86\pm5.40$ & $21.42\pm1.76$ & $6.85\pm0.61$ & $7.37\pm0.15$ & $38.30\pm7.72$ & $56.54\pm5.38$ & $46.61\pm7.80$ & $46.31\pm6.57$\\
M6 & local K4 & $11.45\pm0.94$ & $6.96\pm0.76$ & $3.05\pm0.04$ & $5.53\pm0.70$ & $39.50\pm3.52$ & $37.98\pm7.02$ & $39.05\pm5.87$ & $30.11\pm2.60$\\
\bottomrule
\end{tabular}%
}
\end{table}

M12 lies outside the training interval when held out. Its normalized coordinate is 2.2308, so Eq.~\eqref{eq:domain_guard} rejects it before decoding. For auditability, we also evaluate the unguarded polynomial: degree 2 gives sampled distribution errors of 204\%, 454\%, and $1.81\times10^6$\%, while degree 3 gives $1.97\times10^{15}$--$3.09\times10^{16}$\%. Figure~\ref{fig:full_mach} shows the resulting scale separation and a representative density ratio. A polynomial evaluated beyond its normalized training interval can grow rapidly, while positivity constrains only the sign of $f$ and does not bound its integrated moments. The stress test documents the consequence of bypassing the range guard; it is not an admissible prediction.

Replacing the legacy M6 reference moments by the grid-refined fine solution changes the local-K4 aggregate errors by less than 0.21 percentage points. At M12 the largest legacy-to-refined reference change is 6.46\% for stress, small relative to the intentionally unguarded stress-test values. The experiments therefore provide complementary deployment controls: moment-resolved holdouts quantify in-range accuracy, and the deterministic guard blocks queries beyond the trained interval.

\section{Cavity results}

\subsection{Physical and transport fields at Knudsen number 0.075}

\begin{figure}[!tbp]
\centering
\includegraphics[width=\textwidth]{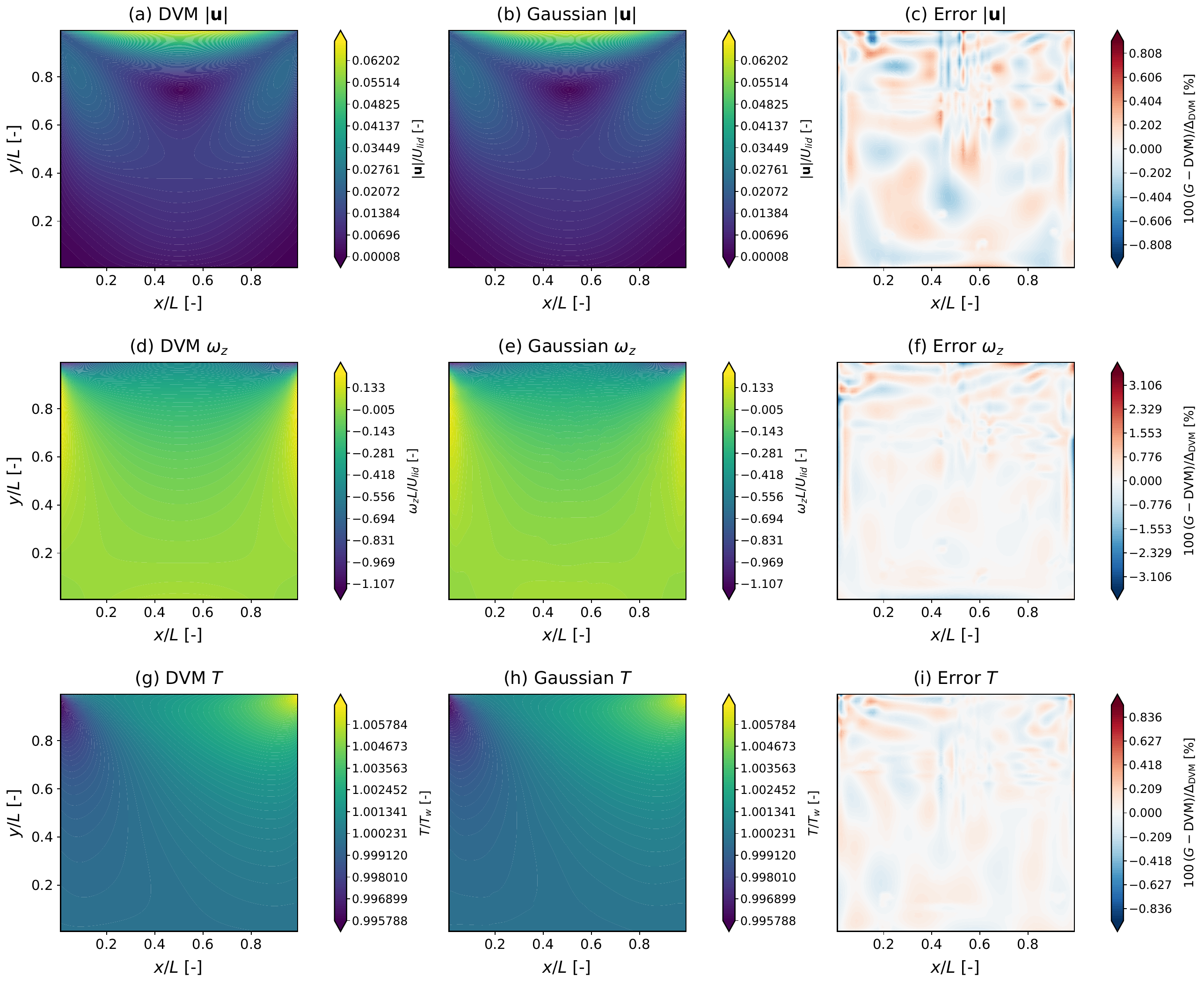}
\caption{Physical cavity fields at $\mathrm{Kn}=0.075$ for the $N=256$ Gaussian map: speed magnitude, vorticity, and temperature. DVM and Gaussian panels use identical color limits within each row. Error panels show the signed percentage of the DVM field range, avoiding singular pointwise percentages near zero.}
\label{fig:cav075physical}
\end{figure}

The cavity representation is assessed first at $\mathrm{Kn}=0.075$. Figure~\ref{fig:cav075physical} uses $N=256$ rather than the lower-error $N=512$ model. The DVM and Gaussian panels place the lid shear layer, primary recirculation, and temperature variation at the same locations on the displayed grid. Vorticity is computed by differentiating reconstructed velocity and is not one of the 20 trained channels; its error map is therefore reported as a separate derivative diagnostic rather than included in the 20-channel maximum. The figure supports comparison of the displayed structures, while the global channel errors are reported numerically below.

\FloatBarrier

\begin{figure}[!tbp]
\centering
\includegraphics[width=\textwidth]{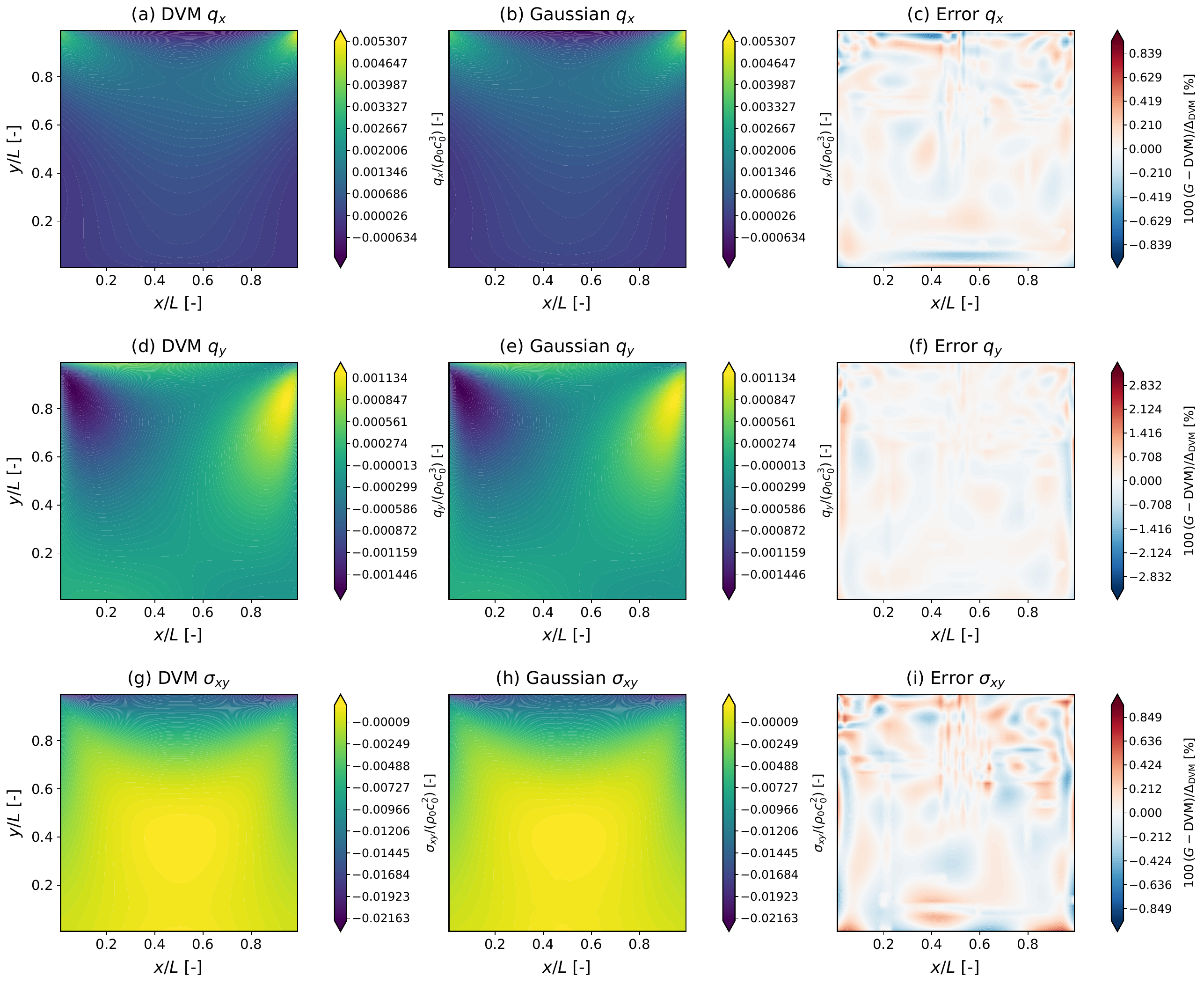}
\caption{Nonequilibrium cavity fields at $\mathrm{Kn}=0.075$ for $N=256$: $q_x$, $q_y$, and $\sigma_{xy}$. The error definition and colorbar conventions are the same as in Fig.~\ref{fig:cav075physical}.}
\label{fig:cav075noneq}
\end{figure}

Heat flux and shear stress provide a closer analogue of the shock nonequilibrium observables. In Fig.~\ref{fig:cav075noneq}, $q_x$ is concentrated near the moving-lid region, $q_y$ changes sign across the cavity, and $\sigma_{xy}$ contains wall-adjacent and interior structure. The DVM and Gaussian panels show the principal extrema in the same regions. On the $65\times65$ reference grid, the $N=256$ model has maximum error 1.455\% over all 20 channels, occurring in $q_y$; at $N=512$ the maximum is 0.326\%.

\FloatBarrier

\subsection{Higher rarefaction at Knudsen number 1}

\begin{figure}[!tbp]
\centering
\includegraphics[width=\textwidth]{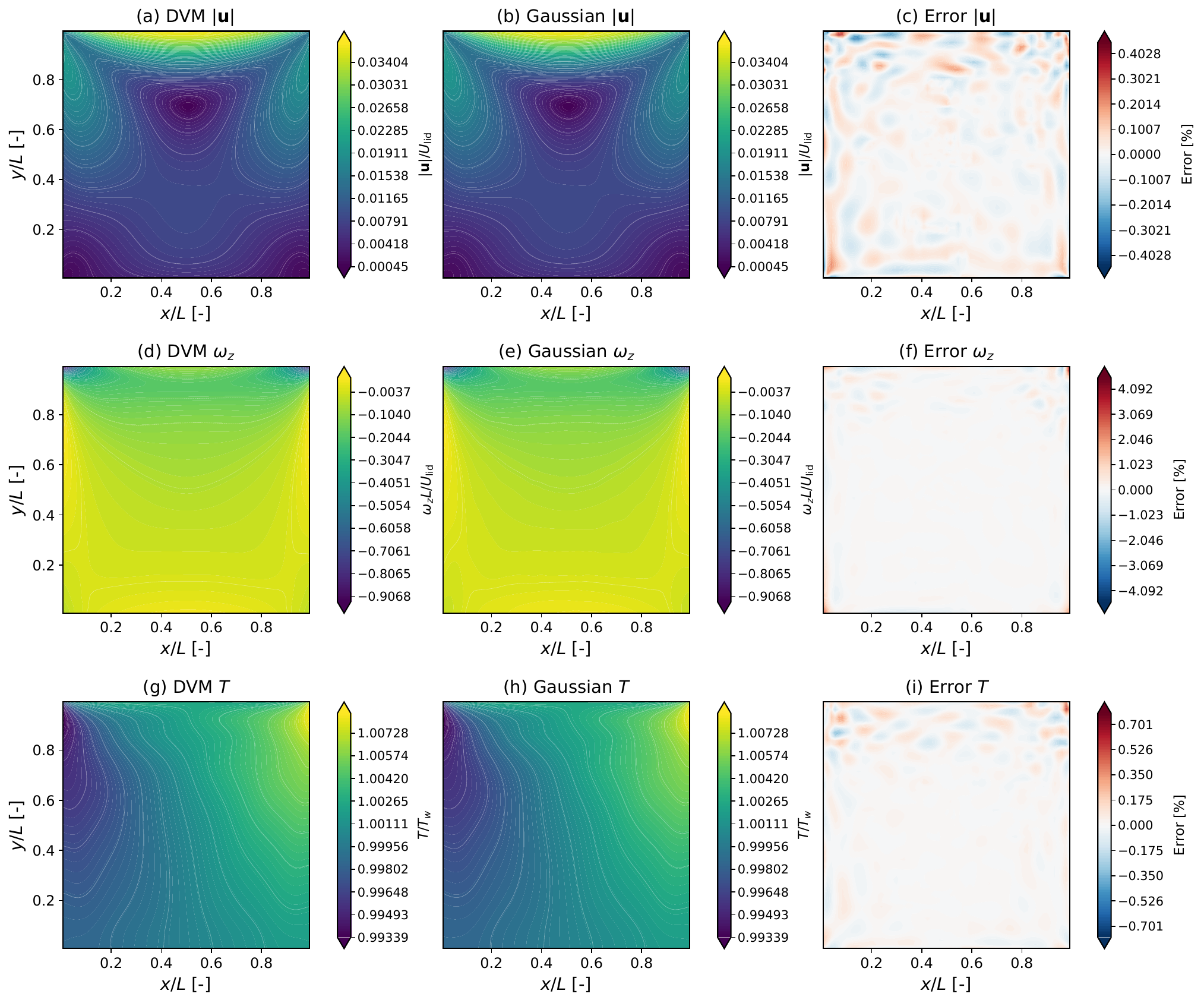}
\caption{Physical cavity fields at $\mathrm{Kn}=1$ for the $N=512$ Gaussian map. DVM and Gaussian fields use identical color limits within each row; the third column reports the signed range-normalized error.}
\label{fig:cav1physical}
\end{figure}

Figure~\ref{fig:cav1physical} repeats the physical-field comparison at $\mathrm{Kn}=1$ using the $N=512$ model. The DVM vorticity differs visibly from the lower-Knudsen case, and the Gaussian panel reproduces its displayed location and sign. As at lower $\mathrm{Kn}$, vorticity is a derived differentiation test rather than a fitted channel.

\FloatBarrier

\begin{figure}[!tbp]
\centering
\includegraphics[width=\textwidth]{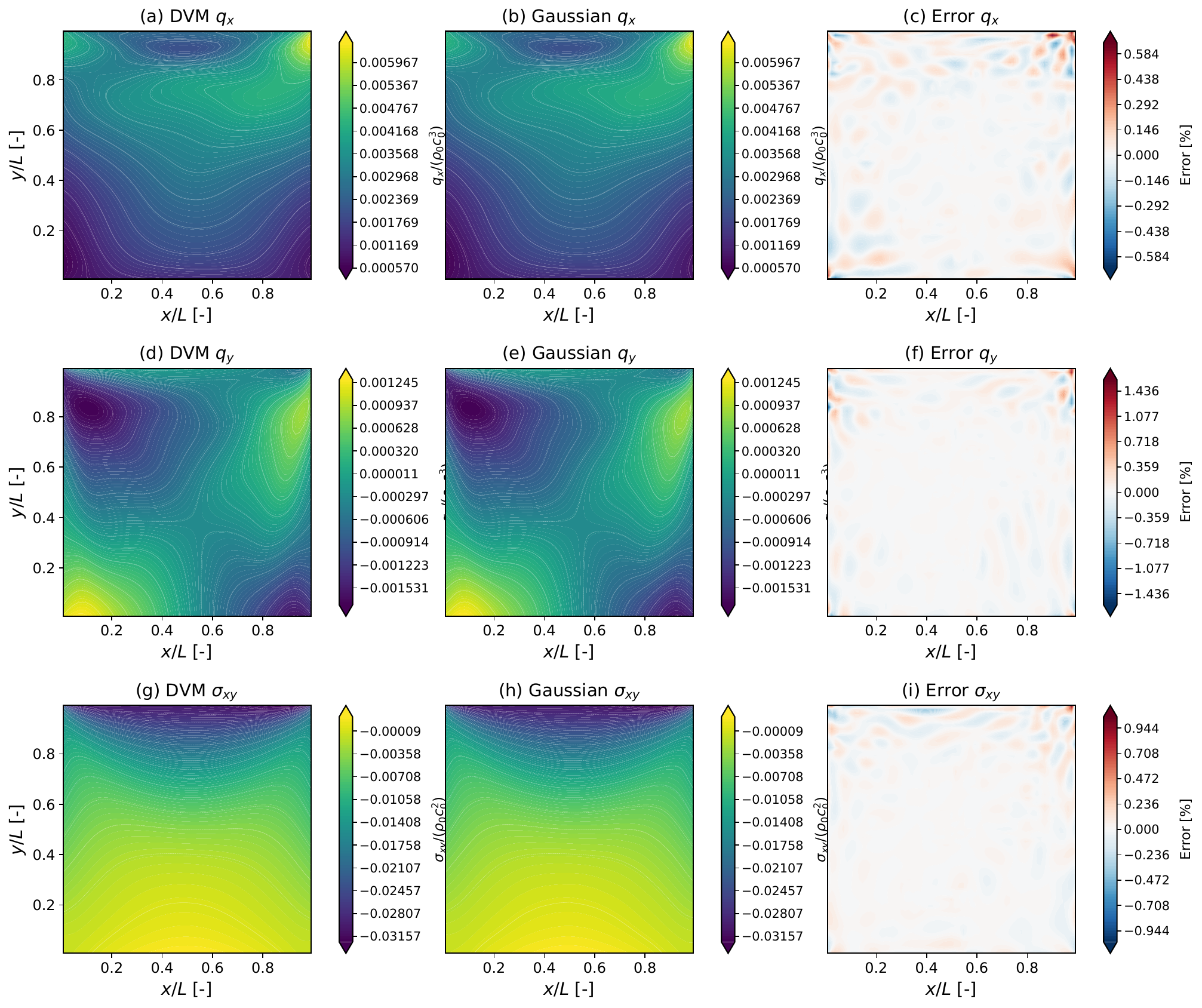}
\caption{Nonequilibrium cavity fields at $\mathrm{Kn}=1$ for $N=512$. The vertical heat flux $q_y$ is the most difficult of the 20 channels, while $q_x$ and $\sigma_{xy}$ remain below 0.2\% relative error.}
\label{fig:cav1noneq}
\end{figure}

Figure~\ref{fig:cav1noneq} reports the corresponding heat-flux and stress fields. Their displayed locations and signs are reproduced by the Gaussian map, and the maximum error over the 20 fitted channels is 0.462\%, again in $q_y$. Together with $\mathrm{Kn}=0.075$, this case tests the shared-support representation under two distinct rarefaction conditions.

\subsection{Compression-fidelity trends}

Table~\ref{tab:cavity_accuracy} compares two storage budgets for both Knudsen numbers. Doubling $N$ reduces the maximum error from 0.995\% to 0.462\% at $\mathrm{Kn}=1$ and from 1.455\% to 0.326\% at $\mathrm{Kn}=0.075$. The fieldwise values in Tables~\ref{tab:full075} and \ref{tab:full1} show that velocity, heat flux, stress, and third-order channels set the larger errors. Directional-temperature, raw fourth-order, and $K_i$ channels contain large near-equilibrium backgrounds and are therefore interpreted separately from the shock $M_{400}^{\mathrm{neq}}$ deviation.

\begin{table}[t]
\caption{Cavity accuracy and nominal compression. Errors are percentages. ``Low-order max'' is the maximum over $\rho,u,v,T$; ``transport max'' is the maximum over $q_x,q_y,\sigma_{xx},\sigma_{yy},\sigma_{xy}$; ``all-field max'' is over all 20 channels.}
\label{tab:cavity_accuracy}
\centering
\begin{tabular}{ccccccc}
\toprule
$\mathrm{Kn}$ & $N$ & parameters & compression & low-order max & transport max & all-field max\\
\midrule
0.075 & 256 & 6164 & 13.71 & 0.624 & 1.455 & 1.455\\
0.075 & 512 & 12308 & 6.87 & 0.170 & 0.326 & 0.326\\
1 & 256 & 6164 & 13.71 & 0.554 & 0.995 & 0.995\\
1 & 512 & 12308 & 6.87 & 0.253 & 0.462 & 0.462\\
\bottomrule
\end{tabular}
\end{table}

The centerline overlays in Fig.~\ref{fig:dvm_centerlines} supplement the global norms by displaying selected profile shapes. Along those centerlines, the $N=512$ Gaussian curves track the DVM reference across the plotted wall layers and sign changes, consistent with the full-grid channelwise norms and contour maps.

\subsection{Matched-budget bilinear and SVD comparisons}

\begin{figure}[!tbp]
\centering
\includegraphics[width=\textwidth]{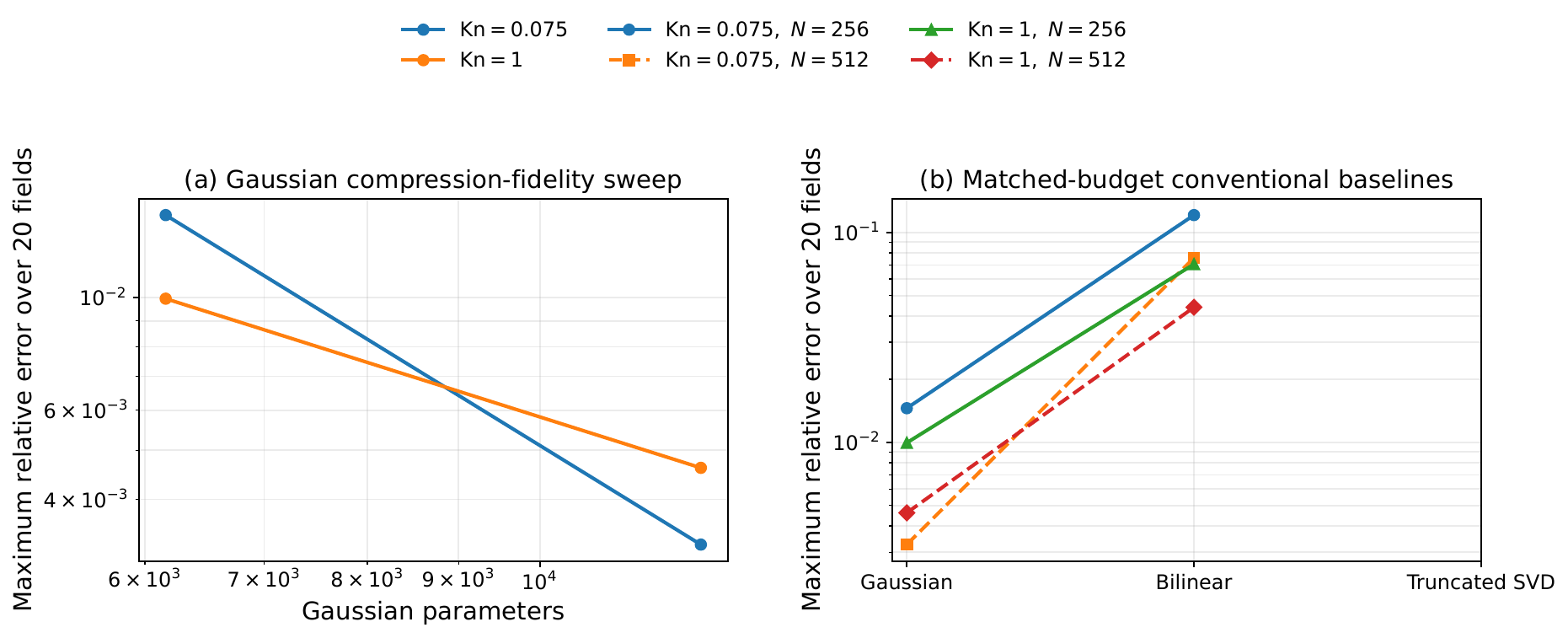}
\caption{Cavity compression and matched-budget comparisons. (a) Maximum 20-field Gaussian error versus parameter count. (b) Gaussian, uniform-grid bilinear, and per-field truncated-SVD errors at the two budgets and Knudsen numbers. Line plots make the budget and method trends explicit.}
\label{fig:cavity_baselines}
\end{figure}

Figure~\ref{fig:cavity_baselines} compares the maximum 20-field error as a function of Gaussian capacity and against two conventional reconstructions; Table~\ref{tab:cavity_baselines} lists the corresponding maxima and coefficient counts. At $N=256$, the Gaussian model is 8.3 times more accurate than matched-grid bilinear interpolation for $\mathrm{Kn}=0.075$ and 7.1 times more accurate for $\mathrm{Kn}=1$. At $N=512$, the factors are 23.2 and 9.5. The truncated SVD is also less accurate: its maximum errors range from 4.06\% to 23.95\%, compared with 0.326--1.455\% for Gaussians. Because grid dimensions and SVD rank are discrete, exact coefficient matching is unavailable: the tested baselines use 6--15\% fewer coefficients than the Gaussian models. The exact counts are reported in Table~\ref{tab:cavity_baselines}, and the resulting budget mismatch is small relative to the observed error gaps.

\begin{figure}[!tbp]
\centering
\includegraphics[width=\textwidth]{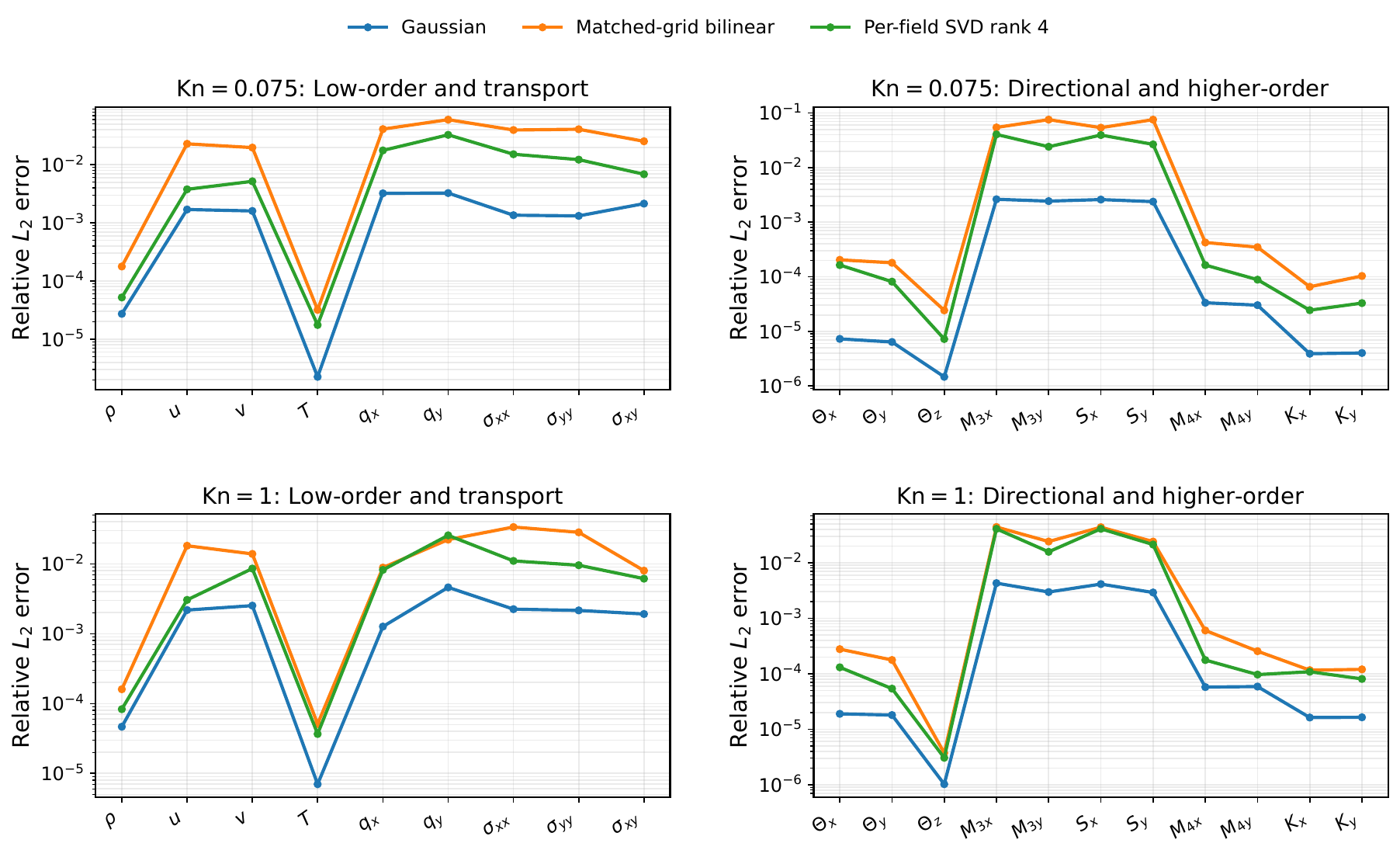}
\caption{Fieldwise relative errors at the 12,308-parameter Gaussian budget, separated into low-order/transport and directional/higher-order channel families for $\mathrm{Kn}=0.075$ and 1. The matched bilinear grid stores 11,520 values and the rank-4 per-field SVD stores 10,500 coefficients. Gaussian errors are consistently lower on the transport and third-order channels that set the maximum.}
\label{fig:cavity_fieldwise}
\end{figure}

The fieldwise comparison in Fig.~\ref{fig:cavity_fieldwise} identifies which channels set the maxima. In the tested bilinear reconstructions, the largest errors occur in third-order channels; the rank-4 SVD also has its largest errors in third-order channels. The Gaussian errors are lower for every displayed transport and third-order channel at the $N=512$ budget, linking the aggregate improvement to the observables most sensitive to localized nonequilibrium structure.

\begin{table}[t]
\caption{Maximum relative error over all 20 cavity fields. Stored values include all coefficients required by each reconstruction.}
\label{tab:cavity_baselines}
\centering
\begin{tabular}{cccccc}
\toprule
$\mathrm{Kn}$ & Gaussian $N$ & Gaussian & bilinear & truncated SVD & baseline storage\\
\midrule
0.075 & 256 & 1.455\% & 12.098\% & 19.514\% & 5780 / 5260\\
0.075 & 512 & 0.326\% & 7.565\% & 4.056\% & 11520 / 10500\\
1 & 256 & 0.995\% & 7.062\% & 23.950\% & 5780 / 5260\\
1 & 512 & 0.462\% & 4.401\% & 4.081\% & 11520 / 10500\\
\bottomrule
\end{tabular}
\end{table}

\section{Computational cost and practical use}

Table~\ref{tab:cost} reports the measured M5 training times retained in the reproducibility tables. The diagonal seed-1234 run required 2.56 h on an NVIDIA Tesla M40 24-GB GPU, while the three correlated runs averaged $2.68\pm0.59$ h. These measurements characterize the tested implementation only; no training-speed advantage is claimed. The 4608 diagonal parameters represent $2.261\times10^8$ M5 phase-space values, giving the nominal coefficient ratio in Table~\ref{tab:shock_baseline}. Moment evaluation still requires the DVM velocity quadrature.

\begin{table}[t]
\caption{Measured M5 training cost. The correlated value is mean $\pm$ sample standard deviation over three seeds; the diagonal timing is the representative seed-1234 run used in Fig.~\ref{fig:m5}.}
\label{tab:cost}
\centering
\small
\setlength{\tabcolsep}{4pt}
\begin{tabular}{@{}lccc@{}}
\toprule
Model & runs & parameters & time (h)\\
\midrule
Diagonal $N=512$ & 1 & 4608 & 2.56\\
$x$--$\xi_x$, $N=512$ & 3 & 5120 & $2.68\pm0.59$\\
\bottomrule
\end{tabular}
\end{table}

The conditional parameter counts are given in Table~\ref{tab:conditional_protocol}: a 15,360-parameter map stores 61,440 float32 bytes and a 20,480-parameter map stores 81,920 bytes, regardless of whether the basis is global or local. These counts describe representation storage, not the cost of generating the eight BGK-DVM training states.

The cavity has a different storage profile. Its raw 20-field state contains 84,500 values, giving the coefficient ratios in Table~\ref{tab:cavity_accuracy}. At $N=512$, the model uses 12,308 parameters; Table~\ref{tab:cavity_baselines} shows maximum-error reductions by factors of 9.5--23.2 relative to the tested bilinear baselines and 8.8--12.4 relative to the tested rank-4 SVD baselines.

The common Gaussian support strategy produces two outputs matched to the archived state. The shock realization is a positive kinetic object from which additional quadrature moments can be computed after training; the cavity realization uses the same localized-support logic to couple 20 signed wall-transport observables. Their shared role is compact, continuous recovery of transport information, while the output parameterization reflects whether the archive contains a VDF or a prescribed field set.

\section{Discussion}

The conceptual unity of the two benchmarks lies in localized Gaussian support rather than identical model outputs. In the shock problem, kernels tile the joint spatial--velocity domain and positivity makes the decoded object a distribution from which moments are regenerated. In the cavity problem, kernels tile the physical domain and share their support across 20 signed outputs. Both constructions replace a dense sampled state by a continuous coefficient set, and both are evaluated by transport fidelity under explicit coefficient accounting. The output parameterization is therefore adapted to the available kinetic state while the representation principle and evaluation criteria remain the same.

\subsection{What controls kinetic fidelity?}

Within the controlled M3 ablation in Table~\ref{tab:m3_ablation}, changing from sampled moment supervision to direct log-density training lowers every reported error, including a reduction of the stress error by more than a factor of ten. Within the three-seed M5 comparison in Table~\ref{tab:m5_seeds}, the mean differences between diagonal and $x$--$\xi_x$ models are smaller than one sample standard deviation for each nonequilibrium observable. Log-density training and the simpler diagonal form are therefore the supported choices for the reported configurations. A broader ranking of objectives and covariance structures would require a larger designed comparison.

This result complements the continuum Gaussian-field study of Shenoy and Frankel, where anisotropy is evaluated for derivative-sensitive vortical structures.\cite{ShenoyFrankel2026} Their spatial analysis and the present phase-space analysis examine different decoded objects, but both ask how Gaussian geometry affects physically sensitive observables. Here the decisive tests are moment reconstruction by DVM quadrature and consistency across random seeds.

\subsection{Why matched storage matters}

A reconstruction error becomes operationally meaningful only when storage is accounted for. At the M5 4608-coefficient budget, all tested multilinear four-dimensional grids give transport and higher-moment errors above 87\%, whereas the diagonal Gaussian errors are 1.12--1.80\% (Table~\ref{tab:shock_baseline}). In the cavity, Tables~\ref{tab:cavity_baselines}, \ref{tab:full075}, and \ref{tab:full1} show smaller Gaussian maximum errors by factors of 7.1--24.1 across the tested bilinear and SVD controls. These matched-budget comparisons isolate the advantage of localized continuous support over the specific grid and low-rank controls used here; sparse-grid, tensor, and other structured representations remain useful future benchmarks.

The nominal compression ratios are transparent coefficient counts. A production archive can additionally exploit symmetry, nonuniform grids, lossless compression, or reduced distributions, while Gaussian parameters require finite precision and metadata. Applying one accounting rule to every method makes the present comparisons reproducible and separates representational structure from file-format choices.

\subsection{Spatial continuity and parameter-range guard}

The interleaved-station test in Table~\ref{tab:holdout} supports spatial interpolation within the tested M5 state: training and holdout errors differ by at most 1.3\% of the corresponding training error in the shock layer. Complete-state exclusion defines a stricter boundary. The M3 global holdouts have 23--42\% transport and higher-moment errors, and the M6 global D3 model has 42.86\% distribution error. The four-knot M6 map lowers that distribution error to 11.45\% without increasing coefficient count, but its transport and higher-moment errors remain 30--40\%. Thus local conditioning improves the matched-budget M6 distribution fit, while the moment audit identifies the accuracy still required for deployment.

The M12 audit in Fig.~\ref{fig:full_mach} separates an in-range query from an out-of-range query. Once the upper training boundary is M8, the normalized M12 coordinate is 2.2308 and Eq.~\eqref{eq:domain_guard} rejects it. The intentionally unguarded calculation shows that bounded centers, positive widths, and positive amplitudes guarantee sign but do not impose Rankine--Hugoniot states or global moment bounds outside the fitted interval. Range screening and moment accuracy are distinct: the former rejects M12, while the latter must still be measured for in-range states such as M3 and M6.

\subsection{Relation to kinetic solvers and physics-informed learning}

The demonstrated uses are compact archival of fitted DVM states, continuous evaluation between tested spatial stations, and quadrature generation of moment hierarchies from a positive decoded distribution. These capabilities can support kinetic-state databases, initialization, and training-data generation, provided an in-range query also passes a whole-state acceptance test with tolerances chosen for the intended observables. Equation~\eqref{eq:domain_guard} supplies the complementary range check.

The same Gaussian ansatz could be embedded in a physics-informed kinetic solver by treating its parameters as outputs of a network and evaluating streaming, collision, conservation, and boundary residuals through quadrature. Positivity would then be built into the distribution parameterization. The current study supplies the representation and identifies the conservation, acceptance, and domain controls needed for that extension.

\subsection{Scope and extensions}

The present benchmarks combine localization in high-dimensional phase space for normal shocks with multi-field wall transport in square cavities. Natural extensions include conservative local coefficient maps, curved external geometries, wall-bounded phase-space models, and residual-trained Gaussian kinetic solvers. Heat flux remains the most seed-sensitive observable, making adaptive tail sampling and transport-aware placement of centers particularly promising. The reported errors quantify representation fidelity to the specified BGK- and Shakhov-DVM targets; transfer to other collision models, geometries, or boundary conditions requires separate validation.

\section{Conclusions}

For separately fitted Mach-3 and Mach-5 shocks, Figs.~\ref{fig:m3} and \ref{fig:m5} and Tables~\ref{tab:m3_ablation} and \ref{tab:m5_seeds} show that 512 diagonal kernels reconstruct $\rho$, $u_x$, and $T$ below 0.25\% and give approximately 1--2\% errors in heat flux, stress, and third- and fourth-order nonequilibrium moments. The M3 objective ablation favors log-density training among the tested objectives, while the three-seed M5 data do not distinguish diagonal and correlated kernels. Table~\ref{tab:holdout} shows that interleaved M5 stations retain nearly the same errors as training stations, and Tables~\ref{tab:shock_baseline} and \ref{tab:grid_layouts} show that the 4608-parameter Gaussian model is substantially more accurate than the tested multilinear four-dimensional grids at matched storage.

At $\mathrm{Kn}=0.075$ and 1, Table~\ref{tab:cavity_accuracy} gives maximum errors of 0.326\% and 0.462\% over the 20 fitted fields for the shared 512-kernel cavity map. Across both budgets, Table~\ref{tab:cavity_baselines} shows that the Gaussian maximum error is lower than the tested bilinear and per-field SVD errors by factors of 7.1--24.1. Figures~\ref{fig:dvm_centerlines}, \ref{fig:cavity_dvm_high_moments}, \ref{fig:cav075physical}, \ref{fig:cav075noneq}, \ref{fig:cav1physical}, and \ref{fig:cav1noneq} document the distinct changes in low-order, heat-flux, stress, and higher-moment fields between the two Knudsen numbers.

The nine-Mach audit in Fig.~\ref{fig:full_mach} and Table~\ref{tab:mach_holdout} separates training-state reconstruction, in-range holdout accuracy, and out-of-range rejection. Global degree-2 and degree-3 models fit their training shocks with mean sampled distribution errors of 1.87\% and 1.27\%, but neither the M3 nor M6 global holdouts is high fidelity. A four-knot M6 map reduces the distribution error from $42.86\pm5.40\%$ to $11.45\pm0.94\%$ at the same 20,480-parameter budget, while 30--40\% transport and higher-moment errors remain. A held-out M12 query is rejected because its normalized coordinate 2.2308 lies beyond the M8 training boundary. Table~\ref{tab:shock_grid_cert} shows that independent M6 and M12 DVM grid studies keep reference discrepancies below 0.12\% and 0.97\%, respectively, which is much smaller than the reported whole-state errors.

Across both flow classes, the central result is a common localized-support principle adapted to the available state: a positive, moment-generative phase-space representation for shocks and a shared-support signed field map for cavities. Matched coefficient accounting connects this principle to storage, while transport moments, complete-state holdouts, and the range guard connect it to use. The present claims concern compact reconstruction of the specified BGK- and Shakhov-DVM targets; broader physical and geometric transfer remains a validation task for conservative and residual-informed extensions.

\appendix
\section{Full cavity fieldwise errors at the high-accuracy budget}

Tables~\ref{tab:full075} and \ref{tab:full1} report all 20 channel errors for the Gaussian, bilinear, and truncated-SVD reconstructions at the $N=512$ Gaussian budget. Values are percentages. The full tables show that heat-flux and third-order channels control the maximum. Raw directional-temperature, fourth-order, and $K_i$ channels carry substantial near-equilibrium backgrounds, so their small relative errors are interpreted separately from the shock nonequilibrium deviations.

\begin{table}[t]
\caption{Fieldwise relative errors in percent for $\mathrm{Kn}=0.075$ at the $N=512$ Gaussian budget. The bilinear and SVD baselines store 11,520 and 10,500 coefficients, respectively, compared with 12,308 Gaussian parameters.}
\label{tab:full075}
\centering
\scriptsize
\begin{tabular}{lrrrrrrrrrr}
\toprule
Method & $\rho$ & $u$ & $v$ & $T$ & $q_x$ & $q_y$ & $\Theta_x$ & $\Theta_y$ & $\Theta_z$ & $\sigma_{xx}$\\
\midrule
Gaussian & 0.0027 & 0.170 & 0.160 & 0.00022 & 0.322 & 0.326 & 0.00073 & 0.00064 & 0.00015 & 0.135\\
Bilinear & 0.018 & 2.288 & 1.973 & 0.0032 & 4.097 & 5.964 & 0.020 & 0.018 & 0.0024 & 3.957\\
SVD rank 4 & 0.0052 & 0.378 & 0.518 & 0.0018 & 1.763 & 3.270 & 0.016 & 0.0081 & 0.00072 & 1.516\\
\bottomrule
\end{tabular}
\vspace{4pt}

\begin{tabular}{lrrrrrrrrrr}
\toprule
Method & $\sigma_{yy}$ & $\sigma_{xy}$ & $M_{3x}$ & $M_{3y}$ & $S_x$ & $S_y$ & $M_{4x}$ & $M_{4y}$ & $K_x$ & $K_y$\\
\midrule
Gaussian & 0.132 & 0.214 & 0.263 & 0.243 & 0.260 & 0.238 & 0.0033 & 0.0030 & 0.00039 & 0.00040\\
Bilinear & 4.078 & 2.530 & 5.445 & 7.565 & 5.393 & 7.562 & 0.042 & 0.035 & 0.0066 & 0.010\\
SVD rank 4 & 1.224 & 0.687 & 4.056 & 2.413 & 3.947 & 2.666 & 0.016 & 0.0088 & 0.0024 & 0.0033\\
\bottomrule
\end{tabular}
\end{table}

\begin{table}[t]
\caption{Fieldwise relative errors in percent for $\mathrm{Kn}=1$ at the $N=512$ Gaussian budget. Storage counts are the same as in Table~\ref{tab:full075}.}
\label{tab:full1}
\centering
\scriptsize
\begin{tabular}{lrrrrrrrrrr}
\toprule
Method & $\rho$ & $u$ & $v$ & $T$ & $q_x$ & $q_y$ & $\Theta_x$ & $\Theta_y$ & $\Theta_z$ & $\sigma_{xx}$\\
\midrule
Gaussian & 0.0047 & 0.218 & 0.253 & 0.00070 & 0.127 & 0.462 & 0.0019 & 0.0018 & 0.00010 & 0.225\\
Bilinear & 0.016 & 1.822 & 1.391 & 0.0051 & 0.886 & 2.228 & 0.028 & 0.018 & 0.00037 & 3.377\\
SVD rank 4 & 0.0083 & 0.305 & 0.858 & 0.0037 & 0.821 & 2.567 & 0.013 & 0.0054 & 0.00031 & 1.104\\
\bottomrule
\end{tabular}
\vspace{4pt}

\begin{tabular}{lrrrrrrrrrr}
\toprule
Method & $\sigma_{yy}$ & $\sigma_{xy}$ & $M_{3x}$ & $M_{3y}$ & $S_x$ & $S_y$ & $M_{4x}$ & $M_{4y}$ & $K_x$ & $K_y$\\
\midrule
Gaussian & 0.216 & 0.192 & 0.429 & 0.297 & 0.412 & 0.291 & 0.0058 & 0.0059 & 0.0016 & 0.0016\\
Bilinear & 2.835 & 0.801 & 4.401 & 2.411 & 4.367 & 2.406 & 0.060 & 0.025 & 0.012 & 0.012\\
SVD rank 4 & 0.958 & 0.615 & 4.071 & 1.573 & 4.081 & 2.125 & 0.018 & 0.0097 & 0.011 & 0.0081\\
\bottomrule
\end{tabular}
\end{table}

\FloatBarrier

\section{Checkpoint audit}

Table~\ref{tab:checkpoint} reports the best-versus-terminal audit for all three correlated seeds and all three diagonal seeds. Only one of the six minimum-loss checkpoints has a lower heat-flux error than its terminal counterpart. Thus the smallest recorded minibatch loss does not define a uniformly superior kinetic checkpoint. A fixed terminal rule avoids selecting a model with the reported evaluation observable.

\begin{table}[t]
\caption{M5 heat-flux error for the checkpoint with the smallest recorded training loss (best) and the terminal checkpoint at 80,000 steps. Values are percentages.}
\label{tab:checkpoint}
\centering
\begin{tabular}{lrrr}
\toprule
Model & seed & best & terminal\\
\midrule
$x$--$\xi_x$ & 1234 & 1.564 & 1.243\\
$x$--$\xi_x$ & 2026 & 5.442 & 3.595\\
$x$--$\xi_x$ & 3407 & 1.141 & 1.232\\
Diagonal & 1234 & 1.139 & 1.125\\
Diagonal & 2026 & 2.713 & 2.639\\
Diagonal & 3407 & 1.626 & 1.456\\
\bottomrule
\end{tabular}
\end{table}

\FloatBarrier

\section{Reproducibility details}

The accompanying archive contains configurations for the principal shock and cavity representation cases, stored prediction profiles, seed and checkpoint tables, matched-storage baselines, and figure-generation scripts. It additionally contains all 18 global full-Mach evaluations, all six local-M6 evaluations, aggregate machine-readable tables, grid-refined M6 and M12 moment profiles, DVM accounting records, and checksums. Shock coordinates are normalized by the midpoint and half-width of each stored coordinate. The bounded center and scale parameterizations, initialization values, optimizer settings, sampling fractions, checkpoint interval, and seeds stated in Sec.~III are reproduced in machine-readable form. The adaptive-$x$ grid in Table~\ref{tab:grid_layouts} places nodes by weighted quantiles of a score formed from $|\partial_x\rho|$, $|q_x|$, and $|\sigma_{xx}|$. These records make the principal conditional, matched-storage, and grid-refinement results traceable to explicit stored artifacts; two legacy single-state diagnostics are retained as reported recomputations rather than used as independent headline evidence.

\section*{Declaration of competing interest}
The author declares that he has no known competing financial interests or personal relationships that could have appeared to influence the work reported in this paper.

\section*{CRediT authorship contribution statement}
Ehsan Roohi: Conceptualization, Methodology, Software, Validation, Formal analysis, Investigation, Data curation, Visualization, Writing -- original draft, Writing -- review and editing.

\section*{Funding}
This research did not receive any specific grant from funding agencies in the public, commercial, or not-for-profit sectors.

\section*{Data availability}
The accompanying reproducibility archive contains processed data, trained-model outputs, machine-readable configurations and tables, global and local Mach-conditioning records, DVM grid-refinement profiles, baseline results, and figure-generation scripts for the principal reported comparisons. The complete DVM phase-space arrays and model checkpoints are available from the corresponding author upon reasonable request because of their size.

\section*{Declaration of generative artificial intelligence and assisted technologies in the manuscript preparation process}
During preparation of this work, the author used OpenAI ChatGPT, including Deep Research, and Anthropic Claude for language editing, code organization and debugging, manuscript structuring, and critical review before submitting the paper to the journal. The author independently reran and verified the computations, checked the numerical results and references, reviewed and edited all content, and takes full responsibility for the work.

\end{document}